\documentclass[prb,superscriptaddress,showkeys,twocolumn,amsmath,amssymb]{revtex4-2}
\usepackage{graphicx}
\usepackage{comment}
\usepackage[colorlinks=true,allcolors=blue]{hyperref}
\usepackage{braket}
\usepackage[normalem]{ulem}
\usepackage{soul}
\usepackage{simpler-wick}

\begin{document}

\title{Limitations of Caldeira-Leggett model for description of phase transitions in superconducting circuits.}
\author{O. Kashuba and R.-P. Riwar}
\affiliation{Peter Gr\"unberg Institute, Theoretical Nanoelectronics, Forschungszentrum J\"ulich, D-52425 J\"ulich, Germany}

\begin{abstract}
The inherent complexity of system-bath interactions often requires making critical approximations, which we here show to have a radical influence on the renormalization group flow and the resulting phase diagram. Specifically, for the Caldeira-Leggett model Schmid and Bulgadaev (SB) predicted a phase transition, whose experimental verification in resistive superconducting circuits is currently hotly debated. For normal metal and Josephson junction array resistors, we show that the mapping to Caldeira-Leggett is only exact when applying approximations which decompactify the superconducting phase. We show that there exist treatments that retain phase compactness, which immediately lead to a phase diagram depending on four instead of two parameters. While we still find an SB-like transition in the transmon regime, the critical parameter is controlled exclusively by the capacitive coupling. In contrast, the Cooper pair box maps to the anisotropic Kondo model, where a pseudoferromagnetic phase is not allowed for regular electrostatic interactions.

\end{abstract}

\maketitle

\section{Introduction}
A realistic depiction of any quantum system necessitates precise models to incorporate dissipation.
For practical calculations, however, straightforward tracing out of bath degrees of freedom is often possible for non-interacting excitations only.
For quantum circuits, there are two common types of resistors, either normal metals or superconducting transmission lines, both of which can under certain conditions be mapped to noninteracting bosons (via the Luttinger liquid paradigm~\cite{Haldane_1981,Pham_2000}, or by neglecting inductive nonlinearities~\cite{Altland_Simons_book,Bruder_2005,Koch_2009,Manucharyan_2009,Catelani_2011}, respectively).
These mappings, however, break charge quantization---which is justified only when charge is defined with a fuzzy spatial resolution~\cite{Rajaraman_1982,Haldane_1981} and when conserving the size of the Hilbert space~\cite{Riwar_2021,Koliofoti_2023}. Generically, if an operator $A$ has integer eigenvalues, the moment generating function $\langle e^{i\lambda A}\rangle$ is $2\pi$-periodic in the counting field $\lambda$~\cite{Aristov_1998,Gutman2010,Ivanov_2013,Ivanov_2016,kashuba2023counting}.  Crucially, the very same moment generating function structures are widely present in renormalization group schemes~\cite{Leggett_1987,Giamarchi_1988}, where $\lambda$ now parametrizes the coupling and enters in the critical exponent, and $A$ represents the environment operator the system couples to~\footnote{To see explicit moment generating function structures pertinent to the RG flow, consider, e.g., Eqs.~(3.18)-(3.20) in Ref.~\cite{Leggett_1987}, or Appendix~A in Ref.~\cite{Giamarchi_1988}.}.
We are unaware of existing works that address the impact of such spectral properties of $A$ for renormalization group flow equations -- as most charge-charge interactions indeed involve a finite length scale. In this work, we caution that depending on how the interaction is described, the spectral properties of $A$ may change radically, and thus significantly impact the RG flow.

Moreover, since charge (in units of the number of Cooper pairs $N$) and the superconducting phase ($\phi$) form a pair of canonically conjugate observables, $[N,\phi]=i$ \cite{Devoret_1997,Burkard_2004,Vool_2017,Riwar_2022}, quantization of the former imposes compactness of the latter~\cite{Peierls_book,Likharev_1985}, such that the gate-induced offset charge $N\rightarrow N+N_g$ is \textit{not} an irrelevant gauge term~\cite{Cottet2002,Koch_2007}. Indeed, while the superconducting phase must be fundamentally $2\pi$-periodic at each point in time and space~\cite{Altland_Simons_book}, for certain circuit elements, like linear inductive shunts, it is justified to apply a low-energy approximation that effectively breaks phase compactness, where $N_g$ is gauged away. For the fluxonium, for instance, this approximation implies that one never returns to the actual charge qubit for infinite inductance~\cite{Koch_2009}. A similar statement is true for resistive shunts, where it was likewise criticized very recently~\cite{Murani_2020} that circuits with infinite resistance (which should essentially mean no dissipation) do not seem to return to the uncoupled version. This seemingly strange observation is best understood along the lines of Refs.~\cite{Loss_1991,Mullen_1993}, which argue that a noncompact phase arises due to entanglement of the inductor or environment with the number of windings in $\phi$. Inspired by this understanding, Ref.~\cite{Koliofoti_2023} has recently developed a revised description of quantum phase slip junctions, where (among other results) it becomes directly evident, that already weak quantum phase slips immediately restore $N_g$ as a physically relevant parameter.

In this work, we combine the above fundamental aspects regarding charge quantization, and show that they lead to a significantly modified understanding of dissipative phase transitions in superconducting circuits.
For circuits coupled to a generic resistor, the Caldeira-Leggett model remains the most extensively studied and utilized to date. When applied to a circuit featuring a single Josephson junction (JJ), it is formulated as~\cite{Caldeira_1981} 
\begin{equation}\label{eq_CL}
\begin{split}
H \!=\! E_CN^2 \!-\! E_J\cos\left(\phi\right) \!+\! \sum_j\!\left[\frac{2e^2}{C_j}N_j^2 \!+\! \frac{\left(\phi_j \!-\! \lambda_j\phi\right)^2}{8e^2 L_j}\right]
\end{split}
\end{equation}
where the charging energy of the superconducting island is $E_C=2e^2/C$, while $[N_j,\phi_{j^\prime}]=i\delta_{jj^\prime}$ describe a fictitious ensemble of LC resonators (representing the resistive element). The coupling to the phase difference $\phi$ across the junction (which we refer to as ``electric'' coupling~\footnote{In circuits, there exist the following main coupling types: charge-charge coupling due to capacitive interactions, current-current coupling, which is of inductive nature, and exchange of electrons across an interface. While capacitive interactions couple to the charge $N$, both inductive coupling and electron transfer usually couple to the phase $\phi$, such that one could in principle cast both of them under the umbrella term of ``inductive'' coupling. In our work, it will be of importance that the resistor allows for the transfer of charges, which is why we make explicitly this distinction.}) via the parameter $\lambda_j$ results in a resistance $R$ experienced by the circuit~\cite{Ingold:1992aa}.

Schmid~\cite{Schmid_1983} and Bulgadaev~\cite{Bulgadaev_1984} discovered that this model undergoes a superconductor-insulator transition (hereafter referred to as the SB transition) at $R_Q/R=1$ ($R_Q=\pi/2e^2$ denotes the resistance quantum in terms of the Cooper pair charge $2e$, with $\hbar=1$). Although this finding has been theoretically substantiated in numerous subsequent studies~\cite{Aslangul_1985,Guinea_1985,Schoen_1990,Herrero_2002,Kimura_2004,Werner_2005,Lukyanov_2007}, both experimental validation~\cite{Murani_2020,Hakonen_2021,Murani_2021,kuzmin2023observation,houzet2023microwave,burshtein2023inelastic} as well as certain aspects of the theoretical treatment~\cite{Masuki_2022,Sepulcre_2023,Masuki_2023,giacomelli2023emergent,altimiras2023absence} are to this day highly controversial.
We take a different direction: instead of exploring or revisiting the phenomenology predicted by Eq.~\eqref{eq_CL}, we critically reexamine its universal applicability, either for a circuit coupled to a piece of normal metal (the resistor used in the experiments of Refs.~\cite{Murani_2020,subero2023bolometric}), or to a transmission line made of Josephson junction arrays~\cite{kuzmin2023observation}.

As a dissipative generalization of the fluxonium~\cite{Catelani_2011}, the Caldeira-Leggett Hamiltonian likewise makes sense for a noncompact phase only. For superconductor-normal metal heterostructures, it has been shown~\cite{Schoen_1990} that there exists a mapping to Eq.~\eqref{eq_CL}, exploiting a bosonization akin to a Luttinger liquid~\cite{Haldane_1981,Belitz_1994,Rollbuehler_2001}.
We instead derive a low-energy Hamiltonian which integrates out the boson (photon) instead of the fermion degrees of freedom, and identify crucial differences with respect to Eq.~\eqref{eq_CL}.
In our model, the phase remains compact even in presence of the resistive coupling.
$N_g$, thus, enters as a tuning parameter, such that the transmon ($E_J>E_C$)~\cite{Koch_2007,Schreier_2008} and the Cooper pair box regimes ($E_J<E_C$)~\cite{Bouchiat_1998,Nakamura_1999} exhibit completely different phase diagrams.
In addition, and contrary to the Caldeira-Leggett model, our treatment differentiates between two very different couplings to the resistor: a capacitive coupling and a direct electric coupling.
Importantly, only the capacitive interaction couples to a fuzzy, and thus nonquantized, charge, whereas the electric contact exchanges integer numbers of Cooper pairs [at least withing the quite robust Bardeen's tunneling approximation in Eq.~\eqref{eq_H_SN}]. 
Consequently, while we still find an SB-like transition in the transmon regime, the critical parameter depends \textit{only} on the capacitive coupling, which is in large parts independent of the normal metal resistivity---the parameter modified in Ref.~\cite{Murani_2020}. 

In the Cooper pair box regime, we find that the system is described by an anisotropic Kondo model~\cite{Anderson_1970,Tsvelick_1983}, where the two quasidegenerate charge states play the role of a pseudospin. This mapping enables us to predict the existence of a pseudoferromagnetic phase with suppressed Cooper-pair exchange. But this phase transition can only occur for a coupling with a negative capacitance. Electrostatics with ordinary (e.g., vacuum or dielectric) permittivity cannot give rise to negative capacitances, such that this transition seems forbidden. We note though that (partial) negative capacitances are a proven principle with ferroelectric materials~\cite{Landauer_1976,Catalan_2015,Hoffmann_2020}, or can be engineered via coupling to a nearby system with special effective screening properties~\cite{Little_1964}, see, e.g.,  attractive interactions in quantum dots via coupling to ``polarizers''~\cite{Hamo_2016,Placke_2018}. Very recently, it was proposed that a coupling to auxiliary transmons can even create capacitive interactions that are quasiperiodic in charge space~\cite{Herrig_2023,herrig2024}.


Finally, we consider the Josephson junction array implementation of a resistor, where we argue that the same picture holds qualitatively. Adopting a path integral language, we show that if we explicitly keep quantum phase slip processes finite and insist on a compact phase, the dissipative part of the action likewise does not yield the Caldeira-Leggett result for the electric contact.

Our updated treatment of the system-bath interaction thus showcases the importance of charge quantization for dissipative quantum phase transitions -- either in the form of nontrivial gauge aspects, or due to the special role of integer-valued observables in renormalization schemes.
As we outline, there remains one potentially important detail concerning non-adiabatic contributions due to a coupling to higher energy states. While higher levels have to some degree been included into analytic renormalization schemes of the Kondo model~\cite{Pustilnik_2001}, to the best of our knowledge there exist no treatments are tailored to our problem.
We, therefore, expect that more precise predictions will necessitate numerical methods. Thus, we note that the fermionic model (normal metal resistor) is readily amenable to a full numeric calculation of the phase diagram by means of numeric RG methods~\cite{Bulla_2008}, envisaged for a follow-up research effort.


\section{Results}

\subsection{The normal metal resistor}
Instead of bosonizing the Hamiltonian describing the superconducting-normal metal (SN) interface~\cite{Schoen_1990}, we stick to a fermionic description which accounts for two types of interactions: a direct electric contact via the proximity effect and a capacitive coupling between the Cooper pair charge on the island. The Hamiltonian is (Appendix~\ref{apx:proximity})
\begin{equation}\label{eq_H_SN}
\begin{split}
&H=E_C \left(N+\sum_{kk^\prime\sigma}\lambda_{kk^\prime}^z c_{k\sigma}^\dagger c_{k^\prime\sigma}\right)^2-E_J\cos(\phi)\\ 
&\!+\!\sum_{kk^\prime}\left( e^{-i \phi} \lambda_{kk^\prime}^\perp c^\dagger_{k\uparrow}c^\dagger_{k^\prime \downarrow}\!+\!\text{h.c.}\right)\!+\!\sum_{k\sigma}\epsilon_k c_{k\sigma}^\dagger c_{k\sigma}\!+\!\ldots,
\end{split}
\end{equation}
where $\lambda^z$ and $\lambda^\perp$ parametrize the capacitive coupling and electric contact, respectively. The notation is obviously borrowed from Kondo model physics, interpreting the quantized charge $N$ as a type of generalized spin. The normal metal electrons with momentum $k$ and spin $\sigma$ are annihilated (created) with the fermionic operators $c_{k\sigma}^{(\dagger)}$.
The normal metal shall further be in a grand canonical ensemble with a well-defined chemical potential (which we simply set to be zero for reference).

In Appendix~\ref{apx:proximity}, we explicitly show how to arrive at the different interactions in $H$ (the capacitive coupling $\sim \lambda^z$ and the electric contact $\sim \lambda^\perp$). The capacitive coupling coefficient $\sim\lambda^z$ can be conveniently expressed in terms of a participation ratio of the capacitances between S and N and to ground (Appendix~\ref{apx:electrostatic}) (a treatment that is valid in a pseudo-electrostatic regime, taking the speed of light to infinity). Specifically for the electric contact $\sim\lambda^\perp$, it is important to stress that the Hamiltonian in above form is valid at energies sufficiently below the superconducting gap $\Delta$, \textit{independent} of the quality of the SN interface (i.e., both for weak and strong tunneling across SN). For concreteness, we deploy in Appendix~\ref{apx:proximity} a perturbative Schrieffer-Wolff approximation, which is guaranteed to be a good approximation in the limit of weak tunneling across SN. We stress though, that the above form of $H$ remains valid also for strong coupling: in that regime, the induced pairing ($\sim\lambda^\perp$) has to be determined self-consistently, which changes the value of the pairing strength, but still leads to the exact same pairing term, expressing the exchange of Cooper pairs at the SN interface, (see also Ref.~\cite{Hosseinkhani_2018} and references therein). Note in particular that even for good SN contact, the induced pairing amplitude in the N region is always smaller than $\Delta$ (as any proximity-induced pairing can naturally never exceed the original superconducting gap). On the other hand, the coupling $\lambda^{\perp}$ does \emph{not} have to be small with respect to the characteristic system energies like $E_{C}$ or $E_{J}$.
We do perform the expansion over it later [in Eqs.~\eqref{eq_R_SN} and~\eqref{eq_ES_coupling}], though we imply a small parameter $\nu\lambda^{\perp}$, where $\nu$ is the density of states of the fermionic bath.

The dots ($\ldots$) at the end of Eq.~\eqref{eq_H_SN} indicate that the normal metal may by all means include a whole host of other physics, such as disorder, electron-electron~\cite{capacitive_footnote}, and electron-phonon interactions.
Many-body interactions will of course (as already stated in the introduction) render it difficult to make quantitative predictions. However, we note that we will be able to draw a number of very concrete qualitative conclusions that do not necessitate solving the bath Hamiltonian. For concrete quantitative predictions, we assume that screening allows for a description in terms of non-interacting excitations.
Let us further emphasize that static disorder can (at least formally) be included in the single-particle picture, by simply diagonalizing the single-particle Hamiltonian of the uncoupled normal metal with a given impurity potential configuration. Consequently, the level index $k$ in general no longer refers to a ballistically propagating wave, but to the set of eigenstates of the Hamiltonian in presence of disorder, leading to an updated $k$-dependence of both energies $\epsilon_k$ as well as the coefficients $\lambda^z,\lambda^\perp$. Of course, our approach (which formally does not require averaging over ensembles of impurity configurations) has the disadvantage that it is no longer straightforward to cast the resulting resistance felt by the circuit in terms of the diffusive properties of the metal (e.g., mean free path). Indeed, diffusion is most conveniently captured by diagrammatic techniques involving artificial averaging over impurities which conveniently introduces a relaxation process, see Ref.~\cite{Gonzalez_2021} for an instructive review). The crucial advantage of our approach is however, that the indirect inclusion of a bulk resistance allows us to make statements that do not depend on the details of the metal bulk. In particular, while the coupling $\sim \lambda^\perp$ seems to nominally represent only the contact resistance, we can take it to effectively include the intrinsic normal metal resistivity. From a phenomenological point of view, we expect this approach to be sufficient, as in practice the actual resistance felt by the circuit is often obtained as a fitting parameter when experimentally characterizing the device under investigation.



Let us now point out the main differences between the models of Eqs.~\eqref{eq_H_SN} and~\eqref{eq_CL}.
As foreshadowed, Eq.~\eqref{eq_H_SN} allows for a compact representation of the phase $\phi$ even in the presence of the bath (that is, we can impose $2\pi$-periodicity on the wave functions of the full Hamiltonian). Consequently, and contrary to Eq.~\eqref{eq_CL}, the system is no longer insensitive to offset charges $N\rightarrow N+N_g$, and the phase diagram depends on $N_g$. Since the $N_g$-sensitivity of the qubit depends very strongly on the ratio $E_J/E_C$~\cite{Cottet2002}, we will discuss the different regimes separately.

Phase compactness plays a second important role. We cast Eqs.~\eqref{eq_CL} and~\eqref{eq_H_SN} into a more comparable form by transforming the couplings to the phase $\phi$ into an effective capacitive coupling. Applying respective unitary transformations $U^{(\text{CL})}=\prod_j e^{i\lambda_j N_j\phi}$ and $U^{(\text{SN})}=\prod_{q\sigma} e^{ic_{q\sigma}^\dagger c_{q\sigma}\phi/2}$, we get,
\begin{equation}\label{eq_HX}
    H^{(X)}=E_C\left[N+N_\text{env}^{(X)}\right]^2-E_J\cos(\phi)+H_\text{env}^{(X)}\ ,
\end{equation}
with $X=\text{CL},\text{SN}$ for either the Caldeira-Leggett model~\cite{Leggett1984,Ankerhold2007}, or the fermionic model. In this representation, the interaction term universally appears inside the charging energy $\sim E_C$ as a bath-induced shift of the charge, the respective charges being $N_\text{env}^{(\text{CL})}=\sum_j\lambda_jN_j$ and
\begin{align}
\label{eq_NF}
N_\text{env}^{(\text{SN})}&=\underbrace{\sum_{kk^\prime\sigma}\lambda_{kk^\prime}^z c_{k\sigma}^\dagger c_{k^\prime\sigma}}_{N_\text{loc}}+\underbrace{\frac{1}{2}\sum_{k\sigma}c_{k\sigma}^\dagger c_{k\sigma}}_{N_\text{tot}}\ .
\end{align}
Note that the latter terms are equivalent to the exponents in the unitary transformation.
The (uncoupled) environment Hamiltonians read
\begin{align}
\label{eq_HB}
\! H_\text{env}^{(\text{CL})}&\!=\!\sum_j\left[\frac{2e^2}{C_j}N_j^2+\frac{\phi_j^2}{8e^2 L_j}\right]\\ 
\label{eq_HF}
\! H_\text{env}^{(\text{SN})}&\!=\!\sum_{k\sigma}\epsilon_k c_{k\sigma}^\dagger c_{k\sigma}\!+\!\ldots \!+\! \sum_{kk^\prime}\!\lambda_{kk^\prime}^\perp\left(c^\dagger_{k\uparrow}c^\dagger_{k^\prime \downarrow}\!+\!\text{H.c.}\right).
\end{align}
We note that in this basis choice, the Caldeira-Leggett Hamiltonian is likewise $2\pi$-periodic in $\phi$---which naively seems to render our previous concerns moot.
In particular, for a closed system, a $2\pi$-periodic Hamiltonian and a $2\pi$-periodic $\phi$-space are straightforwardly related, in that the former allows for the qubit basis (the system without the reservoir) to be expressed in terms of a Bloch wave vector $k$, whereas for the latter, Bloch wave vector is fixed to the value $k=N_g$, with $N_g$ being the equivalent of the vector potential along the compact $\phi$~\cite{Ulrich_2016,Koliofoti_2023}.

But, if we cast $\phi$ on the circle for the transformed Caldeira-Leggett model, there is no way to get back to the original representation in Eq.~\eqref{eq_CL}, because the above introduced unitary transformation $U^{(\text{CL})}$ is not periodic in $\phi$ and thus ill-defined for compact $\phi$. That is, even though the Caldeira-Leggett Hamiltonian in the basis choice given in Eq.~\eqref{eq_HX} is $2\pi$-periodic in $\phi$, we need to include all (periodic as well as nonperiodic) Bloch states $k$, in order to guarantee the equivalence of Eqs.~\eqref{eq_CL} and~\eqref{eq_HX}. The Bloch state argument will reappear further below for the Kondo model discussion.

Conversely, the Caldeira-Leggett model, Eq.~\eqref{eq_HX}, with \textit{compact} $\phi$ explicitly includes only a capacitive coupling to the charge (for pure capacitive coupling, see also Ref.~\cite{Kaur_2021}). If the bath is both capacitively \textit{and} electrically coupled, however, only the version with extended $\phi$ is applicable. Here, there is only one meaningful total resistance $R$ that does not depend on the physical origin. In contrast, $\phi$ is always compact in fermionic model, such that the different couplings are distinguishable: while we can transform the electric contact into an effective capacitive coupling (it is up to us how to keep track of the number of exchanged Cooper pairs), we cannot simply transform away the entire capacitive coupling into an effective electric contact. At this step, it is not only the quantized Cooper pair charge (expressed in terms of a compact phase), which is important. The two environment charges in Eq.~\eqref{eq_NF} are of very different nature: $N_\text{loc}$ a local portion of the charge within the normal metal, which partakes in the capacitive coupling to the Cooper pair charge. This coupling occurs in general with a finite spatial resolution (a fuzziness, given by, e.g., fringe effects of the electric fields). Consequently~\cite{Rajaraman_1982,Riwar_2021}, the distribution of eigenvalues of $N_\text{loc}$ can by all means be assumed to be (quasi)continuous.
$N_\text{tot}$ on the other hand corresponds to the total charge on the normal metal (in units of the Cooper pair charge), which changes due to dissipative transport across SN. Within the here considered low-energy approximation, the SN interface has an infinitely small length scale, such that, in the absence of quasiparticle processes (more on that later) $N_\text{tot}$ must be integer quantized. Therefore $N_\text{tot}$, and only $N_\text{tot}$, can be subject to unitary transformations defined on a compact $\phi$.
While both types of couplings contribute to the total resistance, they are nonetheless physically distinguishable---particularly regarding poor man's scaling, as we show below.
In particular, we can in a way distinguish these two resistances as either providing an Ohmic or a non-Ohmic dissipation mechanism in the Caldeira-Leggett sense. In this context we note that Ref.~\cite{Ambegaokar_1982,Ambegaokar_1984} have considered the dissipative impact of Bogoliubov quasiparticles (above the superconducting gap) tunneling across the Josephson junction, and have concluded that under certain conditions, they may likewise not provide a true Ohmic dissipation mechanism. In contrast, the coupling to a normal metal resistor (with a continuum of states below the gap) was up to know regarded as purely Ohmic~\cite{Schoen_1990}.

To proceed, the resistance is computed via the quantum Langevin equation~\cite{Ingold:1992aa} (see Appendix~\ref{apx:ohmic})
\begin{align}
\frac{C}{4e^2}\ddot{\phi} &= -E_{J} \sin(\phi) - J(t) +   \int_{t_0}^{t} dt' Y(t-t')\dot{\phi}(t'),
\end{align}
where $J(t) = i[H_\text{env}(t),N_\text{env}(t)]_{-}$ and dissipation is described by $Y(t-t') = [  N_\text{env}(t') ,J(t)]_{-} $.
Assuming weak time-dependence, we can simplify the last term to $R^{-1}\dot{\phi}(t)$.
The resistance for CL and SN take the form 
\begin{align} \label{eq_R_CL}
R_{Q}/R^{(\text{CL})} &= \pi\sum_{k}|\lambda^{z}_{k}|^{2}\omega_{k}\delta(\omega_{k}),
\\ \label{eq_R_SN}
R_{Q}/R^{(\text{SN})} &= \underbrace{\pi |\lambda^{\perp}(0)|^{2} \nu^{2}}_{R_Q/R_{\perp}} + \underbrace{\pi \nu^{2} \lim_{\omega\to0}(\lambda^{z}(\omega)\omega)^{2}}_{R_Q/R_{z}},
\end{align}
respectively.
Here for the CL model we use the designation $\lambda^z_{k}=\lambda_{k}\sqrt[4]{\frac{C_{k}}{4L_{k}}}$ , and for SN $\lambda^{z/\perp}(\omega)$ defined as $\lambda^{z/\perp}_{kk'} = \lambda^{z/\perp}(\omega_{k}-\omega_{k'})$.
Note that while $U^{(\text{SN})}$ eliminates the explicit coupling to $\phi$, it leads to pairing fluctuations in the metal, which gives rise to the first term in Eq.~\eqref{eq_R_SN}.

\subsection{Transmon regime}

\begin{figure}[h]
\centering
\includegraphics[width=.93\columnwidth]{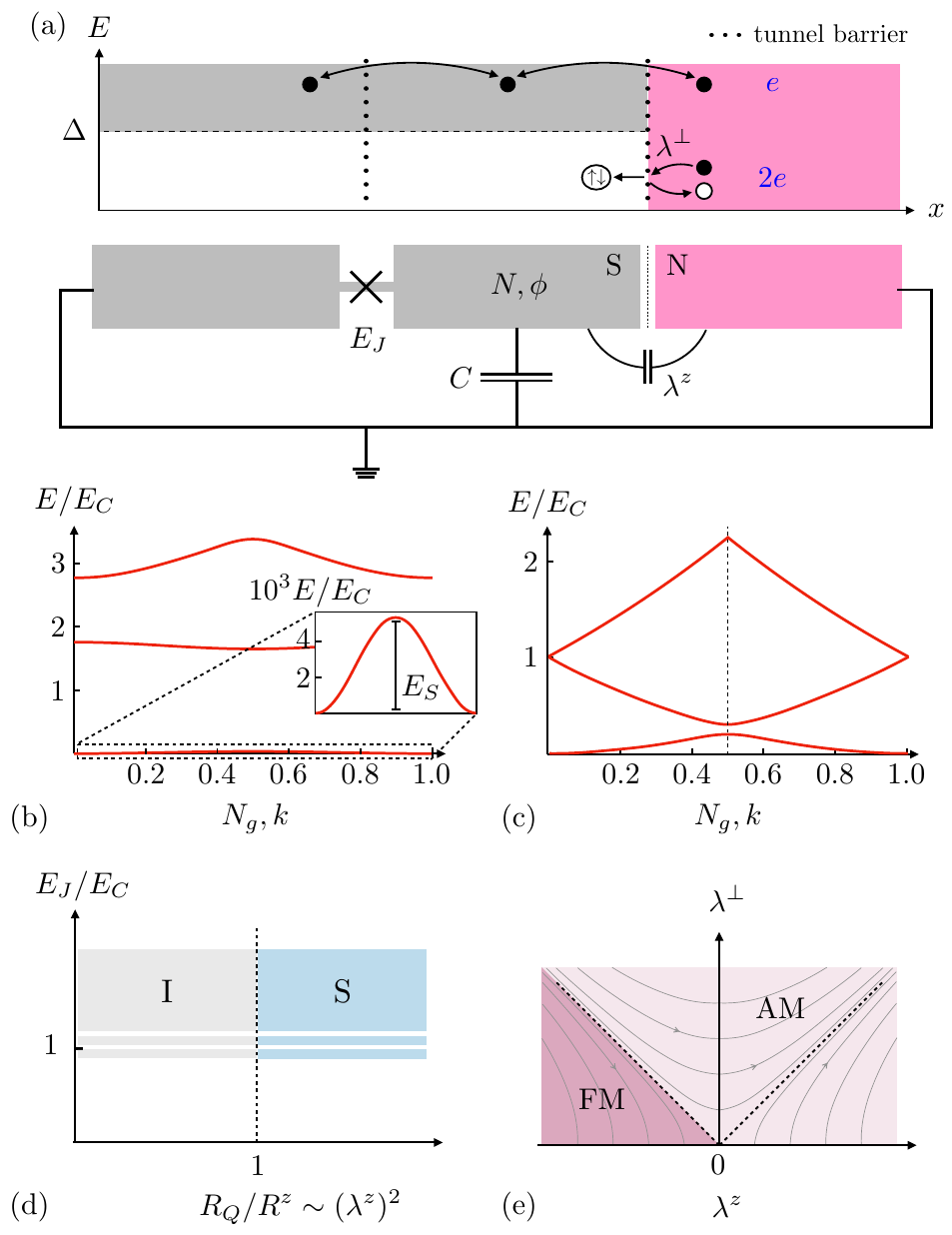}
\caption{An overview of main accomplishments of this work. (a) JJ-based charge qubit coupled to a normal metal with illustration of two main couplings: a capacitive one ($\sim\lambda^z$) between the superconducting island and the normal metal, and an electric contact ($\sim \lambda^\perp$) leading to a lossy Cooper pair exchange via Andreev reflection.
Here, the additional high energy  quasiparticle tunneling processes carrying a single elementary charge $e$ are ignored.
Panels (b)+(c) show the energy regimes $E_J>E_C$ ($E_J=2E_C$) and $E_J<E_C$ ($E_J=0.1E_C$), respectively, as a function of the Bloch wave vector $k$ (for extended $\phi$) or the offset charge $N_g$ (compact $\phi$). (d) In the regime $E_J>E_C$, we still find an insulator-superconductor phase transition (which is insensitive to $N_G$), but with the critical parameter solely determined by the capacitive contribution to the total the resistance [$\sim(\lambda^z)^2$]. 
The more $E_C$ approaches to $E_{J}$, the less reliable is the result, what is indicated by the shading. We did not study the opposite case for general $N_{G}$ (that's why it's left blank), but can cite well-known result for certain values:
(e) For $E_J<E_C$, and tuning $N_g$ close to the charge degeneracy point ($N_g$ close to half-integer), the system may instead exhibit a Kondo-like transition with a pseudoferromagnetic phase. We stress that the full phase diagram depends on four parameters ($E_J/E_C$, $\lambda^z$, $\lambda^\perp$, and $N_g$) and is scarcely studied, which is why we are able to represent it in easy accessible way only by splitting in separate panels (d) and (e).
}
\label{fig_main}
\end{figure}

To illustrate the different roles of both resistance contributions in Eq.~\eqref{eq_R_SN}, we consider $E_J> E_C$~\footnote{The word transmon is commonly associated with very high ratios $E_J/E_C$ (of order $\sim 50$ or higher). We here use the term a little more flexibly, and mean simply that $E_J$ shall be sufficiently large compared to $E_C$, such that $\sqrt{E_J E_C}>E_S$.}. Here the low-energy behavior of the charge qubit reduces to a simple quantum phase slip process, leading to the system-bath Hamiltonian
\begin{equation}\label{eq_H_ES}
    H^{(X)}\approx - E_S \cos(2\pi[N+N_\text{env}^{(X)}])+H_\text{env}^{(X)}\ ,
\end{equation}
where $E_S=16[E_C E_J^3/(8\pi^2)]^{1/4}e^{-\sqrt{32E_J/E_C}}$ and regime $E_S\ll\sqrt{E_C E_J}$ is easy to reach~\cite{Cottet2002}. In accordance with the above discussion, we have two possible types of quantized charges, either the charge on the superconducting island (compact $\phi$), or the possibility of integer-valued contributions in $N_\text{env}^{(X)}$.
For extended $\phi$, we replace the island charge operator $N$ with the Bloch wave vector $k$, representing the fluxonium basis for infinite inductance~\cite{Koch_2009,Catelani_2011}. The resulting energy $-E_S\cos(2\pi k)$ represents quantum tunneling between different minima in the $\cos(\phi)$ potential. For compact $\phi$, we replace $k\rightarrow N_g$, providing the charge dispersion of the transmon ground state, see also inset in Fig.~\ref{fig_main}(b).

To estimate the renormalization of the $E_{S}$ due to the fluctuations of the $N_\text{env}$ we resort to the adiabatic approximation implying that the bath is effectively slow comparing to the superconducting island dynamics~\cite{Leggett_1987,Rymarz2021}.
We stress that this approximation scheme is exact in the limit where $\sqrt{E_J E_C}\gg E_S$ (note that it is always possible to choose the values of $E_J$ and $E_C$ such that $E_S$ remains constant, and the energy gap $\sqrt{E_J E_C}$ is sent to infinity), though the weakening of this constraint introduces only minor quantitative changes to the general statements, see also the discussion below.

Integrating out a given high-energy slice of the bath degrees of freedom (from frequency $\Lambda$ to $\Lambda_0>\Lambda$), the coupling in Eq.~\eqref{eq_H_ES} renormalizes the amplitude $E_S$ (see also Ref.~\cite{Caldeira_1981}), 
\begin{equation}\label{eq_ES_coupling}
    E_S\rightarrow 
    \left<e^{i2\pi N_\text{env}^{(X)}}\right>E_S\ ,
\end{equation}
where $\langle\ldots\rangle \equiv \text{tr}\left[\ldots\right]_{\{\Lambda,\Lambda_{0}\}}$.
This procedure is also referred to as adiabatic renormalization~\cite{Leggett_1987}. For the Caldeira-Leggett model, we get the standard renormalized quantum phase slip amplitude
\begin{equation}\label{eq_ES_ren_CL}
    E_S\rightarrow e^{-R_Q\int_\Lambda^{\Lambda_0}d\omega\frac{Y(\omega)}{\omega}}E_S\ ,
\end{equation}
where $Y(\omega)=\frac{\pi}{2}\sum_j\frac{\lambda_j^2}{L_j}\delta(\omega-\omega_j)$ (with $\omega_j=1/\sqrt{L_jC_j}$) is the abovementioned admittance of the dissipative bath. For an Ohmic bath, $Y(\omega)=1/R^\text{(CL)}$, we arrive at
\begin{equation}\label{eq_dissipative_phase_transition}
    E_S\rightarrow \left(\frac{\Lambda}{\Lambda_{0}}\right)^{\frac{R_{Q}}{R^\text{(CL)}}}E_{S}\ ,
\end{equation}
with the dissipative phase transition marked at the point where the critical exponent $R_Q/R^\text{(X)}=1$, in accordance with Refs.~\cite{Schmid_1983,Bulgadaev_1984}.
Here, we would like to stress that despite obvious equivalence of the capacitive and electric couplings demonstrated in Eq.~\eqref{eq_HX}, the implementation of the physical capacitive ohmic coupling is much more sophisticated than the electric one~\cite{Maile2018,Maile2022}.

For the fermionic model, we similarly integrate out a finite frequency slice from $\Lambda$ to $\Lambda_0$ of the metal's electron degrees of freedom. Due to $[N_\text{loc},N_\text{tot}]=0$ and  the eigenvalues of $N_\text{tot}$ being integer, since the Hamiltonian changes the number of electrons by two, we find $e^{i2\pi N_\text{tot}}=1$, such that $\left<e^{2\pi i (N_\text{loc}+N_\text{tot})}\right>=\left<e^{2\pi iN_\text{loc}}\right>$.
Hence, here the coupling to $\phi$ does \textit{not} directly contribute to the renormalization of $E_S$. Of course, the finite proximity effect term $\sim\lambda^\perp$ will have an indirect influence on $N_\text{loc}$, but this is a higher order effect. In leading order, we can take the trace with respect to the unperturbed normal metal state. This is our first central result: while both capacitive coupling and electric contact contribute to the total resistance, Eq.~\eqref{eq_R_SN}, only the capacitive contribution $\sim \lambda^z$ is relevant for the RG and thus for the SB phase transition, see also Fig.~\ref{fig_main}(d).
To summarize, the capacitive coupling essentially provides a noninteger scrambling of the offset charge, and thus leads to a suppression of the observed $E_S$ by destructive interference. The electric contact cannot accomplish the same effect, as it provides only integer-fluctuations at low energies, which leave the $\cos(2\pi N)$ term unchanged.

Let us reiterate, that this finding is a consequence of standard adiabatic renormalization~\cite{Leggett_1987}. A more precise treatment will likely provide a finite contribution of the $\lambda^\perp$-term to the renormalization of $E_S$, e.g., due to coupling to the transmon excited state (which has a charge dispersion with opposite sign).
Namely, even though the integer contribution to the environment charge $N_\text{tot}$ cannot directly suppress $E_S$, the transmon eigenbasis nonetheless depends on the fluctuating $N_\text{tot}$ such that transitions to higher lying states are possible. As already stated, there is a well-defined limit within the theory ($E_S$ finite, while $\sqrt{E_JE_C}$ large), where this effect can be safely neglected, and our finding is precise. For more moderate ratios of $E_J$ and $E_C$, our finding is simply that on the lowest perturbative level of the RG, the Caldeira-Leggett model massively overestimates the contribution of the electric contact. On a similar token, we comment on Bogoliubov quasiparticle excitations, see Fig.~\ref{fig_main}a.
In the course of the low-energy approximation (energies below the superconducting gap $\Delta$), quasiparticles have been discarded from the theory.
These excitations lead to half-integer (instead of integer) fluctuations of $N_\text{tot}$, and thus likewise contribute to a renormalization of $E_S$.
However, at the very least for weak tunneling, these fluctuations are highly suppressed compared to the fluctuations of $N_\text{loc}$, by the fact that the quasiparticles need to traverse the tunnel barriers. Also here, we therefore expect that the renormalization by means of the capacitive coupling is dominant.
The quantitative statements on this issue require the approaches lying beyond Born-Oppenheimer approximation and, thus, also beyond the scope of this paper.
A numerical treatment for arbitrary parameter regimes is envisaged in the future.


Crucially, Eq.~\eqref{eq_ES_coupling} allows to draw a deep connection between the Ohmic nature of the fermionic system and the full-counting statistics (FCS) problem.
For concreteness, consider the SN junction as a parallel plate capacitor.
The charge $N_\text{loc}$ can then be approximated as $\lambda^z_{0} n$, where $n$ is the local charge at the interface, i.e., the charge on the interval of length $l$.
The moment generating function $\langle e^{i\lambda n}\rangle$ with the counting field $\lambda$ is well-known in 1D~\cite{Aristov_1998,Gutman2010,Abanov2011,LeHur2012,Ivanov_2013,Ivanov_2016,kashuba2023counting}, $\langle e^{i\lambda n}\rangle \approx e^{i\lambda \langle n\rangle - \lambda^{2}(\langle n^{2}\rangle - \langle n\rangle^{2})/2} \propto e^{i\lambda \langle n\rangle - \lambda^{2}(K/2\pi^{2}) \ln \langle n\rangle }$, where $K$ is the Luttinger parameter.
The average charge is proportional to the large ratio of the ultraviolet (Fermi momentum) and infrared cutoffs. Importantly, only the second moment leads to a renormalization of $E_{S}$, such that the logarithmic dependence on $\langle n\rangle$ guarantees an Ohmic behavior with respect to the running frequency cutoff $\Lambda$, i.e., the constant exponent in Eq.~\eqref{eq_dissipative_phase_transition}.
Comparing $\lambda^z_{0} n$ with  $N_\text{loc}$ in Eq.~\eqref{eq_NF} allows us to connect the critical exponent of the SN model with the resistive contribution due solely to capacitive coupling, $R_{Q}/R^z$, in Eq.~\eqref{eq_R_SN}, via $\lambda^{z}_{kk'} = \lambda^{z}_{k-k'} $, where $ \lambda^z_{k} \!=\! \lambda^z_{0} \sin(\pi kl/L)/\pi k \!\approx\! \lambda^z_{0} /\nu \omega$.
Moreover, the similar calculation at $\lambda^{z}_{kk'}\to\delta_{kk'}$ shows that $N_\text{tot}$ can be neglected even for  general counting field, see Appendix~\ref{apx:chargecounting}.
Both calculations give the scaling for the Hamiltonian~\eqref{eq_H_ES} as $E_{S}\to E_{S}(\Lambda/\Lambda_{0})^{a |\lambda^z_{0}|^{2}}$, where $a$ is simply a numeric coefficient.

Through simple dimensional analysis it can be seen that the logarithmic dependence on the ultraviolet cutoff persists 2D and 3D metals. Likewise, interactions or disorder [implied in form of $\ldots$ in Eq.~\eqref{eq_H_SN}] do not change this result, indicating the fundamental ohmicity of the considered fermionic system. A potential exception could be due to electrostatic counter terms $\sim N_\text{loc}^2$ in the Hamiltonian~\cite{capacitive_footnote}, Appendix~\ref{apx:chargecounting}, which have recently been shown to lead to a non-Ohmic bath, as seen from the charge island~\cite{Maile2022}. We note however, that Ref.~\cite{Maile2022} considers a bosonic bath, which connects to a normal metal only for 1D, where screening is well-known to be less effective.

\subsection{Cooper-pair-box regime}

For $E_C>E_J$ and $N_g\approx 1/2$, the fermionic model (with compact $\phi$) maps onto the anisotropic Kondo model.
The island's Fock space is then reduced to $N=-1,0$ and can be represented as a local pseudospin, while $\lambda^z$ and $\lambda^\perp$ play the role of generalized parallel and perpendicular Kondo couplings~\cite{Tsvelick_1983} obeying the phase diagram in Fig.~\ref{fig_main}e)~\cite{Anderson_1970}.
Note that the transition to the pseudoferromagnetic phase (where tunneling across the SN interface is suppressed) takes place only for $\lambda^z<0$, i.e., a \textit{negative} capacitive shunt across the SN junction. In a recent work, it was already noted that the sign of the capacitive coupling is important for a purely capacitively coupled charge qubit (where the resistor was modeled as a transmission line)~\cite{Kaur_2021}. We here extend this result to the combined presence of capacitive and electric coupling. It is therefore interesting to note that while this phase transition seems inaccessible for conventional (vacuum or dielectric) electrostatics, the engineering of negative capacitances (which, as mentioned in the introduction is an actively pursued topic in various works~\cite{Little_1964,Landauer_1976,Hamo_2016,Catalan_2015,Hoffmann_2020,Placke_2018,Herrig_2023,herrig2024}) could potentially render it realizable.

The Cooper pair box regime cannot be reached for the Caldeira-Leggett model for the simple reason that $\phi$ is not compact. Here, the tuning parameter $N_g$ is replaced by a Bloch vector $k$. If we then assume the system to be in a Gibbs ensemble state, the Bloch vector will invariably relax to the ground state, which, for the closed system is simply $k\approx 0$. Overall, we thus see the following difference between the phase diagram resulting from Eqs.~\eqref{eq_CL} and~\eqref{eq_H_SN}. For the standard Caldeira-Leggett model, $E_J/E_C$ and the total resistance $R/R_Q$ [Eq.~\eqref{eq_R_CL}] are the only two relevant tuning parameters. For the fermionic model here, we have in total four relevant tuning parameters. In addition to $E_J/E_C$, the outcome depends strongly on $N_g$, and likewise, the two couplings $\lambda^\perp$ and $\lambda^z$ assume very different roles, in both the Cooper-pair box regime, and the transmon regime [where they contribute to two physically distinct resistive contributions, see Eq.~\eqref{eq_R_SN}].

\subsection{Bosons versus fermions}

For the above normal metal resistor, we obtained a Hamiltonian description and resulting phase diagram that is markedly different from the Caldeira-Leggett model. As pointed out in the introduction, a mapping from the SN interface to the Caldeira-Leggett model has been demonstrated~\cite{Schoen_1990} through integrating out fermionic degrees of freedom. When considering the SN interface in a standard bosonized language (see also Appendix~\ref{apx:bosonization}), we can identify the crucial step, where charge quantization gets lost. Qualitatively speaking, charges transported across the interface are no longer integer quantized, as bosonization blurs the charge density operator over a characteristic length scale (for a homogeneous 1D metal, this would be Fermi wave length~\cite{Haldane_1981}). In this final results section, we want to provide an equivalent, but alternative perspective on this aspect in terms of quantum phase slip processes in the bath, respectively entanglement of phase windings between system and bath. We will do so by comparing normal metal resistors with Josephson junction array resistors, where we detect some formal similarities with regards to charge quantization, phase compactness and bosonization.

\begin{figure}
    \centering
    \includegraphics[width=0.9\columnwidth]{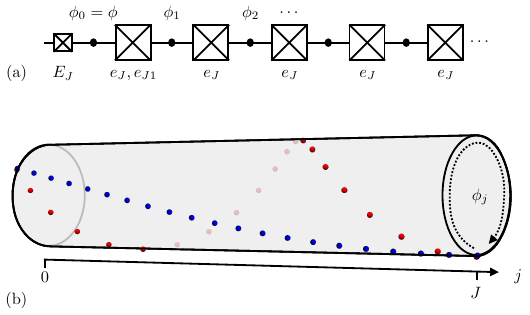}
    \caption{Low-energy treatment of Josephson junction array resistor compatible with phase compactness. (a) The charge island $\phi$ is electrically coupled to the array with nodes $\phi_j$ with $j=1,\ldots,J$ (for notational simplicity we extend the index $j$ to $0$ to include the charge island as $\phi_{j=0}=\phi$). (b) Even when $e_J$ is large, such that the phase differences $\phi_{j+1}-\phi_j$ are small, it is important to consider the configuration space of the total system on a compact manifold (cylinder for the 1D array). An extended $\phi$ is only meaningful when the array has a coupling to ground at the far end $j=J$ (such that the phase $\phi_J$ is strongly coupled to ground, $\phi_J\approx 0$). Here, configurations with different number of windings in the phase profile within the array (blue and red) count the number of times $\phi$ has revolved around the circle (see also arguments in Refs.~\cite{Loss_1991,Mullen_1993,Koliofoti_2023}).  }
    \label{fig_cylinder}
\end{figure}

Let us begin with modeling an environment by means of a Josephson junction array (see also Fig.~\ref{fig_cylinder}a),
\begin{equation}\label{eq_JJ_array}
\begin{split}
    H=E_CN^2-E_J\cos(\phi)-e_J \cos(\phi_1-\phi)\\+e_C\sum_{j=1}^J N_j^2-e_J\sum_{j=1}^{J-1} \cos(\phi_{j+1}-\phi_j)\ ,
\end{split}
\end{equation}
where $j=1,\ldots,J$ indexes the nodes of the array. 
The array can be regarded as a bath in the thermodynamic limit $J\rightarrow \infty$, and assuming (in analogy to the normal metal resistor) it to be in a grand canonical ensemble. We note however, that our line of reasoning below is also valid for finite $J$. There is an interesting alternative case for finite $J$, where in addition the last node is connected back to ground via a term $-e_J\cos(\phi_J)$. The charge qubit and array now form a loop (the actual fluxonium). As we briefly comment at the end of this work, this system works differently, as here, the bath is in a certain regime capable of storing the information of the number of windings that the charge qubit performed in $\phi$-space.

To proceed, while the original array Hamiltonian, Eq.~\eqref{eq_JJ_array}, preserves charge quantization (and thus phase compactness) on each node (both for the central charge island, and the nodes that make up the array), one can make an approximation which is in its essence very similar to the bosonization procedure for fermionic degrees of freedom.
Assuming that the Josephson energies in the array are very large ($e_J\gg E_J,e_C$), we arrive at the Caldeira-Leggett model, Eq.~\eqref{eq_CL}, by approximating the cosine energy-phase relationship of each Josephson junction $\sim e_J$ as $\cos(\delta\phi)\approx 1-\delta\phi^2/2$~\cite{Catelani_2011,Houzet_2019}.
What we have essentially done is to neglect quantum phase slip processes within the array, and extending the phase difference $\delta\phi$ to the real line.
This is by all means a reasonable procedure for computing eigenenergies of the bath (array) Hamiltonian, but, crucially, it again fails to conserve the global properties of the moment generating function $\langle e^{i\lambda N_j}\rangle$, which we have already shown to be relevant for RG, see Eq.~\eqref{eq_ES_coupling}.

In particular, we note that while compactness of the phase \textit{differences} is to some degree irrelevant (since the large $e_J$ guarantees small phase differences) we should still think of the individual phases $\phi_j$ to be compact. That is, an accurate analogy for the system is that of a chain of coupled pendulums [such that the phase profile $\phi_j$ as a function of $j$ can be mapped to a cylinder, see Fig.~\ref{fig_cylinder}(b)]. Consequently, the phase of the charge qubit $\phi$ (and of each array node) is and always has been compact. 

Let us now argue, that this proper consideration of phase compactness is of utmost importance for phase transitions also for JJ arrays.
We again follow the same procedure as for the normal metal resistor, but importantly, we refrain from decompactifying the local phases.
Instead, we take the exact Hamiltonian as given in Eq.~\eqref{eq_JJ_array}, and again perform a unitary transformation, here $U=\prod_j e^{-i N_j \phi}$, to transform the coupling to the $\phi$ into an effective charge-charge coupling, which modifies the charging term again as $N^{2}\to (N+N_\text{tot})^{2}$, where $N_\text{tot}=\sum_j N_j$, and the renormalization prefactor is predicted to be $\langle e^{2\pi i \sum_{j}N_{j}} \rangle$.
Since the canonical charge $N_{j}$ is quantized, all eigenvalues of the charge operators are integer, and the average of the exponent $e^{2\pi i N_{j}}=1$, yet again over \emph{any} ensemble (that is, we can make this conclusion without even solving for the bath Hamiltonian).
This means that in the transmon limit, the renormalization of $E_{S}$ is absent in the same way it already was for the normal metal. 
Note that we may include an actual capacitive coupling to the JJ array, just as we did for the normal metal, which again provides an SB-like phase transition in the transmon regime, as we would here obtain a coupling of the form $N^{2}\to (N+\sum_{j}\alpha_jN_{j})^{2}$, where $\alpha_j$ can assume any real value.

To complete this section, we corroborate the above qualitative discussion of the JJ array by means of more quantitative arguments within the path integral formalism.
Following Ref.~\cite{Altland_Simons_book}, the ohmic bath in the Caldeira-Leggett model contributes with a term $\eta \sum_{\omega_{n}}|\omega_{n}|\phi(\omega_{n})\phi(-\omega_{n})$ to the action (where $\eta=R_Q/R^\text{CL}$, which dominates in between ultraviolet ($\Lambda_{0}$) and infrared ($\Lambda$) cutoffs, and can be written in time domain as 
\begin{equation}
\delta L_\text{bath}^\text{(CL)}\approx\iint d\tau d\tau' K(\tau-\tau')\phi(\tau)\phi(\tau'),
\label{eq:bathKCL}
\end{equation}
where $ K(\tau) =\eta/4\pi^{2} \tau^{2}$.
For small $E_J\ll E_C$, one can perform perturbation theory in $E_J$, and one finds that the Josephson energy is renormalized as $E_{J,\text{eff}} = b^{-1/\eta}E_{J}$, where $b=\Lambda_{0}/\Lambda \gg 1$, Appendix~\ref{apx:chargecounting}. 
In order to see the SB transition, we have to be able to vary $\eta$ on the scales from  $\eta\gg 1$ to $\eta\ll 1$ and still remain in the frequency window, when the dynamics of the field $\phi$ is governed by the dominating bath term $\eta|\omega_{n}|$.
In this regime the average fluctuation of the field is equal to $\phi \sim \sqrt{\eta^{-1}\log b}$.
Thus the contribution of the bath term can be estimated as $\log b$, guaranteeing the strong renormalization.
For the opposite, transmon case $E_{J}\gg E_{C}$, the characteristic energy $E_{S}$ of lowest level coming from the phase slip processes.
In order to estimate the contribution from the bath term~\eqref{eq:bathKCL} we exploit the instanton solution for $\phi(\tau)$, obtaining the value of order of $\sim \eta \log b$, Appendix~\ref{apx:jjarray}, which, in principle, can be large.

To study of the decoupling of the system from the environment and the importance of the closed manifold, we analyze the bath contribution for the JJ array model introduced in Fig.~\ref{fig_cylinder} with a special parameter choice.
We implement a weak coupling to the bath with a Josephson junction with energy $e_{J1}$ (which may be small compared to $e_J$ see Fig.~\ref{fig_cylinder}a), such that its cosine can certainly not be approximated by a parabola. Nonetheless, for the sake of the argument, we assume a parabolic approximation for all other Josephson junctions within the array (treating it as a simple transmission line, with a reservation, see below). Integrating out the phase $\phi_{1}$ we get the bath contribution in the form 
\begin{equation}
\delta L_\text{bath}^\text{(JJa)}\approx \iint d\tau d\tau' K(\tau-\tau')\sin \phi(\tau) \sin \phi(\tau'),
\label{eq:bathKJJa}
\end{equation}
with new $\tilde\eta = \eta (e_{J1,\text{eff}}/\Lambda)^{2}$ (see derivation in Appendix~\ref{apx:bath}).
This term \textit{conserves} phase compactness and for small $\tilde\eta$ the field in delocalized over the whole closed manifold (as long as the bath contribution is dominating), so that it's value can be estimated as $\phi\sim 1$.
Thus the total contribution to the action vanished as $\sim\tilde\eta$ an therefore cannot affect the island's action changing the value of the $E_{J}$.
For the transmon case, the same instantonic solution returns the estimation of the bath term~\eqref{eq:bathKJJa} of order of $\sim\eta$, up to numerical coefficient of order of 1, Appendix~\ref{apx:bath}, and therefore likewise cannot invoke the phase transition.
We additionally have to stipulate the large bath coupling element $e_{J1}\to e_{J}$.
Naively, this case may seem equivalent to the CL model with all cosines expanded.
However, as we have shown above, the phase transition implies a strong delocalization of the phase, and the bath cannot be adequately described by the $\eta |\omega_{n}|$ term.
We hope to resolve this issue in the upcoming numerical treatment.

Overall, while the Caldeira-Leggett model is suitable to describe some traits of Josephson junction array resistors, it does not seem to be adequate to accurately predict phase transitions, at least in the here considered perturbative electric coupling.
Importantly, we expect that this reasoning can be extended for all perturbation orders, while the JJ array model will retain its main property that all fields are entering in the expression as continuous trigonometric functions.
Hence, phase transitions, which in CL model happen due to delocalization of the wave function, cannot overcome the obvious limit of $|(\cos)\sin\phi| < 1$.

We conclude this section with a final remark on Eq.~\eqref{eq:bathKJJa}. Had we formulated the normal metal resistor physics likewise within the path integral language, the electric contribution to resistance ($\sim \lambda_\perp$) would have yielded a similar functional $\phi$-dependence with geometric functions ($\sim \sin,\cos$), and thus would have resulted in the same boundedness of the action. We can thus relate the presence or absence of the phase transition to whether or not a certain physical process can be cast into a true Ohmic form, Eq.~\eqref{eq:bathKCL}, or not. We again refer to Refs.~\cite{Ambegaokar_1982,Ambegaokar_1984}, who noticed the possibility of similar geometric functions of $\phi$ for the related, but distinct case of dissipation via Bogoliubov quasiparticles.

\subsection{Decompactification through entanglement}

In all above cases (normal metal and JJ array resistor) the electric coupling (coupling to $\phi$) could be mapped to a capacitive coupling, where the corresponding unitary transformation yielded a shift of the charge as $N\rightarrow N+N_\text{tot}$ where $N_\text{tot}$ had integer quantized eigenvalues, thus resulting in a coupling that is irrelevant for the RG flow. We here show with the example of the Josephson junction array a possible exception.

Namely, we pick up on the already briefly mentioned variant of the Hamiltonian in Eq.~\eqref{eq_JJ_array}, where we keep $J$ finite and add a Josephson coupling to ground at the end of the chain, $H\rightarrow H-e_J\cos(\phi_J)$. Importantly, note that here, the same unitary as before ($U=\prod_j e^{-iN_j \phi}$) cannot eliminate the coupling to $\phi$, as it simply maps from a coupling $\cos(\phi_1-\phi)$ to a coupling $\cos(\phi_J-\phi)$. There is however a different transformation that can eliminate the coupling to $\phi$ (see also Refs.~\cite{Riwar_2021,Koliofoti_2023}), with $U=\prod_j e^{-i(1-j/[J+1])N_j \phi}$. Note that this transformation is strictly speaking illegal, as it does not conserve compactness of $\phi$. But as shown in Ref.~\cite{Koliofoti_2023}, this procedure can be justified by considering the JJ array part of the Hamiltonian separately from the charge qubit, and assume the phase $\phi$ to be a classically fixed parameter. In the limit of large $e_J$, the phases at both ends of the array are thus pinned close to the values $\phi_1\approx \phi$ and $\phi_J\approx 0$. This pinning however still allows many low-energy solutions, with different number of windings around the cylinder, see Fig.~\ref{fig_cylinder}(b). After appropriately solving the low-energy physics of the JJ array with fixed $\phi$, the system can subsequently reintegrated into the circuit including the charge qubit, and the phase $\phi$ quantized~\cite{Koliofoti_2023}.

In accordance with Refs.~\cite{Loss_1991,Mullen_1993,Koliofoti_2023} it is here reasonable to think of $\phi$ as extended, but only in a very specific sense. Configurations that have the same (compact) value of $\phi$, but different winding number, can be mapped to a different representation wherein $\phi$ is now extended. If we start from a configuration without windings, and let the system evolve (importantly, while neglecting phase slips within the bath), we know that any winding of the system final state contains the precise information of how many times the phase of the charge island $\phi$ has revolved around the circle on which it is defined.

In this regime, the bath perfectly entangles with the number of times the phase $\phi$ revolves around the circle. Consequently, $\phi$ can indeed be regarded as effectively extended (noncompact) and as a consequence, $N_g$ is here an irrelevant gauge term. Note that the intrinsic strain within the array provides in addition a (small) linear inductive energy term $\sim E_L \phi^2$. Moreover, the above unitary creates a different type of environment charge $N\rightarrow N+\sum_j(1-j/[J+1])N_j$, which, crucially, has \textit{no} integer quantized values. Thus, the system where the JJ array and charge qubit form a loop maps in this regime precisely onto the Caldeira-Legget model, in accordance with the findings of Ref.~\cite{Hekking_1997}.

Let us conclude by presenting some final caveats. First of all, the loop description is only valid for a finite size array ($J$ not infinitely large). It must therefore be questioned, whether such an array (at least within the here presented minimal description) actually serves as a true dissipative reservoir. Moreover, as remarked in Ref.~\cite{Koliofoti_2023}, the presence of quantum phase slips (the array may be subject to quantum tunneling between  configurations with different winding number) immediately restores phase compactness, and thus revives $N_g$ as a real physical parameter. In terms of quantum information, the bath now has a finite probability to ``forget'' the number of revolutions of $\phi$ (such that the entanglement is subject to some decay). In the transmon regime ($E_J\gg E_C$), there are thus essentially two quantum phase slip processes, one coming from the charge qubit, the other from the intrinsic tunneling of different winding configurations within the array. These two phase slip amplitudes can be regarded as interfering due to $N_g$ (also known as the cQED version of the Aharonov-Casher effect~\cite{Pop2012,Astafiev_2012}). While the former gets renormalized as per the SB transition, the latter grows with system size (either with $\sim J$ for clean systems, or $\sim \sqrt{J}$ when the charge islands of the array have fluctuating gate offset charges). It was recently predicted~\cite{Houzet_2019} that for very long arrays, quantum phase slips within the array are themselves subject to renormalization according to the RG scheme by Giamarchi and Schulz~\cite{Giamarchi_1988}.

At any rate, the above exception shows that topology enters in more than one way. While the superconducting phase $\phi$ is in principle always compact, closing the JJ array system to a loop effectively allows for its decompactification. With this counterexample, we show that an exact mapping to Caldeira-Leggett is possible if (and only if) the bath is capable of exactly storing the information of the number of times the (compact) phase $\phi$ has revolved around the circle.

\section{Discussion}

We demonstrate that physically distinct dissipation mechanisms (which all lead to a contribution to the resistance) do not have the same impact on the predicted phase diagram of dissipative quantum circuits. By deploying a fermionic model describing capacitive coupling as well as the proximity effect with a normal metal resistor, we find that the mapping to the standard Caleira-Leggett model is not exact. In particular, capacitive coupling turns out to be the dominant factor for phase transitions, rendering the normal metal resistivity (the parameter controlled, e.g., by Ref.~\cite{Murani_2020}) irrelevant. Our results suggest that future experiments could specifically address parts of the device geometry that change the SN capacitance, but not the metal resistance, in order to separate both contributions.
We also identified a Cooper-pair box regime where the dissipative system maps to the anisotropic Kondo model. Also here, capacitive and electric coupling play a distinct role, and in particular, the transition to a pseudoferromagnetic phase is forbidden for regular (repulsive) electrostatic coupling.
We further analyze an alternative realization of a resistor element by means of Josephson junction arrays, and demonstrate that the same conclusions hold qualitatively.
Overall, for both types of resistor realizations, we link the decompactification of the superconducting phase with the approximation of the bath degrees of freedom in terms of non-interacting bosons.
Without this approximation, we show that the electric contact is irrelevant for the RG flow. The only identifiable exception is when the bath is capable of exactly entangling with the number of times the superconducting phase $\phi$ revolves around itself.
Finally, we identify a nontrivial open aspect with respect to nonadiabatic coupling to higher energy states.
In this context, we point out that the fermionic (normal metal) model lends itself to future research endeavors by means of numerical RG methods~\cite{Bulla_2008}, where the currently presented predictions can be refined.

\textit{Acknowledgments}---We warmly thank T.\ Costi, D.\ P.\ DiVincenzo, G.\ Catelani, E.\ König, A.\ Petrescu, \c{C}.\ Girit, C.\ Altimiras, P.\ Joyez and D.\ Maile for fruitful discussions. This work has been funded by the German Federal Ministry of Education and Research within the funding program Photonic Research Germany under the contract number 13N14891.

\appendix

\section{Derivation of proximity effect at SN interface}\label{apx:proximity}

In the main text, we describe the impact of the SN interface at energies below the superconducting gap $\Delta$. Here, we show how the proximity pairing term $\sim \lambda^\perp$ in Eq.~\eqref{eq_H_SN} (due to Andreev reflection) emerges from a Schrieffer-Wolff transformation by eliminating Bogoliubov quasiparticles. We describe the SN interface including excitations at all energies by means of the standard Hamiltonian,
\begin{equation}\label{eq_H_SN_app}
H_{SN}=H_{S}+H_{N}+H_{T}
\end{equation}
with
\begin{align*}
H_{S} & =\sum_{q\sigma}\epsilon_{q}c_{q\sigma}^{\dagger}c_{q\sigma}+\sum_{q}\left(\Delta e^{-i\phi}c_{q\uparrow}^{\dagger}c_{-q\downarrow}^{\dagger}+\text{h.c.}\right)\\
H_{N} & =\sum_{k\sigma}\epsilon_{k}c_{k\sigma}^{\dagger}c_{k\sigma}\\
H_{T} & =\sum_{kq\sigma}\left(t_{kq}c_{q\sigma}^{\dagger}c_{k\sigma}+t_{kq}^{*}c_{k\sigma}^{\dagger}c_{q\sigma}\right).
\end{align*}
We diagonalize the superconducting Hamiltonian,
\begin{equation}
H_{S}=\sum_{q\sigma}\sqrt{\epsilon_{q}^{2}+\Delta^{2}}\gamma_{q\sigma}^{\dagger}\gamma_{q\sigma},
\end{equation}
where
\begin{align}
\gamma_{q\uparrow} & =u_{q}e^{i\frac{\phi}{2}}c_{q\uparrow}+v_{q}e^{-i\frac{\phi}{2}}c_{q\downarrow}^{\dagger}\\
\gamma_{q\downarrow} & =v_{q}e^{-i\frac{\phi}{2}}c_{q\uparrow}^{\dagger}-u_{q}e^{i\frac{\phi}{2}}c_{q\downarrow} \ ,
\end{align}
with
\begin{align}
u_{q} & =\frac{1}{\sqrt{2}}\sqrt{1+\frac{\epsilon_{q}}{\sqrt{\epsilon_{q}^{2}+\Delta^{2}}}}\\
v_{q} & =\frac{1}{\sqrt{2}}\sqrt{1-\frac{\epsilon_{q}}{\sqrt{\epsilon_{q}^{2}+\Delta^{2}}}} \ .
\end{align}
These identities are readily inverted as follows,
\begin{align}
c_{q\uparrow} & =u_{q}e^{-i\frac{\phi}{2}}\gamma_{q\uparrow}+v_{q}e^{-i\frac{\phi}{2}}\gamma_{q\downarrow}^{\dagger}\\
c_{q\downarrow} & =v_{q}e^{-i\frac{\phi}{2}}\gamma_{q\uparrow}^{\dagger}-u_{q}e^{-i\frac{\phi}{2}}\gamma_{q\downarrow} \ .
\end{align}
Note that in the diagonalized $H_{S}$, we have discarded a constant
offset energy, such that the BCS ground state is defined as having energy
$0$.

We now eliminate Bogoliubov quasiparticle excitations by means of
a Schrieffer-Wolff transformation, by projecting onto the BCS ground
state with the projector $P$. We can thus approximate the total Hamiltonian
as
\begin{equation}
H\approx H_{N}+H_{\text{prox}}
\end{equation}
with
\begin{align}
H_{\text{prox}} & =-PH_{T}\left(1-P\right)\frac{1}{H_{S}}H_{T}P\\
 & =\sum_{kk'\sigma}\delta\epsilon_{kk^{\prime}}c_{k\sigma}^{\dagger}c_{k'\sigma}+\sum_{kk'}\left(e^{-i\phi}\lambda_{kk^{\prime}}^{\perp}c_{k\uparrow}^{\dagger}c_{k'\downarrow}^{\dagger}+\text{h.c.}\right)\ ,
\end{align}
with
\begin{align}
\delta\epsilon_{kk^{\prime}} & =-\sum_{q}t_{kq}^{*}t_{k^{\prime}q}\frac{u_{q}^{2}-v_{q}^{2}}{\sqrt{\epsilon_{q}^{2}+\Delta^{2}}}\\
\lambda_{kk^{\prime}}^{\perp} & =-\sum_{q}t_{kq}^{*}t_{k'q}^{*}\frac{2u_{q}v_{q}}{\sqrt{\epsilon_{q}^{2}+\Delta^{2}}}\ .
\end{align}
The $\delta\epsilon$ term is a local impurity potential, which can be included in the normal metal eigenstates by means of diagonalization, thus updating the single electron eigenenergies $\epsilon_k$ and eigenstates. It therefore does not need to be included explicitly in Eq.~\eqref{eq_H_SN}. The $\lambda^\perp$ pairing term on the other hand accounts for the exchange of Cooper pairs at the interface.

\section{Electrostatic interaction at SN interface}\label{apx:electrostatic}

In general, an arbitrary capacitive interaction in the model Hamiltonian, Eq.~\eqref{eq_H_SN} [in the main text], can be represented by two different terms: on-island $N^{2}$ and between island and lead $N \sum_{kk^\prime\sigma}\lambda_{kk^\prime}^z c_{k\sigma}^\dagger c_{k^\prime\sigma}$. One can always rewrite those in form given in Eq.~\eqref{eq_H_SN}, attributing the rest to the interaction-like lead terms.
To estimate the strength of the residual lead terms we address a toy model written in spirit of CL model to make the idea more clear instead of hiding it behind the lengthy mathematical derivations.

\begin{figure}
    \centering
    \includegraphics[width=\columnwidth]{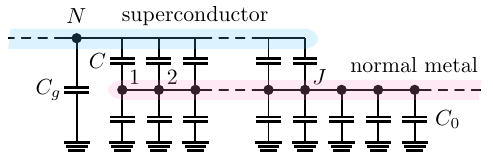}
    \caption{Depiction of the toy model capacitive network describing the electrostatic interaction between superconductor and normal metal. Assuming the Nambu-Goldstone mode within the superconductor to be fast, the superconductor is described with only one lumped charge $N$. The superconductor has a direct capacitance to ground, $C_g$, and a capacitive coupling with the normal metal as $C$. The normal metal itself has a capacitance to ground of its own, $C_0$. }
    \label{fig_network}
\end{figure}

Below we show the derivation of the capacitive interaction $\sim \lambda^z$. This calculation also helps to estimate the order of magnitude of the parameter $\lambda^z$ in terms of a ratio of capacitances. To this end, we solely consider the capacitive network of the SN interface, and consider it as a lumped element network in a parallel plate type of geometry, see Fig.~\ref{fig_network}.

As for the superconducting island, we assume the Nambu-Goldstone mode to be fast, such that Cooper pairs are redistributed quickly. We thus assume that the superconductor can be described by a single pair of charge and phase variables, $N,\phi$. For the normal metal on the other hand we keep the spatial distribution of the charges free, such that there are many normal metal nodes underneath the superconductor, the nodes being enumerated with index $j=1,2,\ldots$. The total number of nodes in the normal metal is irrelevant for the here considered calculation, and is nominally set to infinity. The number of normal metal nodes capacitively coupled to the superconductor is fixed to $J$. Thus, for the capacitive network shown in Fig.~\ref{fig_network}, the Lagrangian is,
\begin{equation}
    L=\frac{C_{g}}{2}\left(\frac{\dot{\phi}}{2e}\right)^{2}+\frac{C}{2}\sum_{j=1}^{J}\left(\frac{\dot{\phi}}{2e}-\frac{\dot{\phi}_{j}}{2e}\right)^{2}+\frac{C_{0}}{2}\sum_{j=1}^{\infty}\left(\frac{\dot{\phi}_{j}}{2e}\right)^{2}\ ,
\end{equation}
where $\dot{\phi}/2e$ is the voltage on the superconductor, whereas $\dot{\phi}_j/2e$ are the voltages on the normal metal nodes.
Through standard Legendre transformation,
\begin{equation}
    H=\dot{\phi}\partial_{\dot{\phi}}L+\sum_j \dot{\phi}_j\partial_{\dot{\phi}_j}L\ ,
\end{equation}
we arrive at the Hamiltonian,
\begin{equation}
\begin{split}
    H=\frac{4e^{2}}{2}\frac{C+C_{0}}{JCC_{0}+\left[C+C_{0}\right]C_{g}}\left[N+\frac{C}{C+C_{0}}\sum_{j=1}^{J}N_{j}\right]^{2}\\+\frac{4e^{2}}{2}\frac{1}{C_{0}}\sum_{j=1}^{\infty}N_{j}^{2}-\frac{4e^{2}}{2}\frac{C}{C_{0}\left[C+C_{0}\right]}\sum_{j=1}^{J}N_{j}^{2}\ .
\end{split}
\end{equation}
The first line has exactly the same shape as the capacitive interaction term in Eq.~\eqref{eq_H_SN} in the main text, when identifying $N_\text{loc}=C/(C+C_0)\sum_j N_j$. The second line provides a simplified account of the electron-electron interactions inside the normal metal. The first term $\sim 1/C_0$ is the ordinary many-body interaction of the uncoupled metal. The second term $\sim C$ indicates that the finite capacitive coupling with the superconductor effectively reduces the interactions within the part of the normal underneath the superconductor (i.e., for $j$ from $1$ to $J$). We refer to this correction as an electrostatic counter term. For $C\ll C_0$ this correction can be discarded. For the opposite case of $C \gg C_0$, this correction may be important. As pointed out in the main text, based on the results of Ref.~\cite{Maile2022} we believe that such corrections could potentially change the nature of the metallic reservoir from Ohmic to non-Ohmic. However, we expect this effect to be present only in a true 1D system, where screening within the metal is weaker.

\section{Computing admittance from quantum Langevin equation}\label{apx:ohmic}

In the main text we provide the formulas for the respective resistances for the Caldeira-Leggett model, Eq.~\eqref{eq_R_CL}, and for the fermionic model, Eq.~\eqref{eq_R_SN}. In this section, we show how these equations are derived from the quantum Langevin equation (similar to Ref.~\cite{Ingold:1992aa}).
We begin with the Hamilton equations for the Hamiltonian in Eq.~\eqref{eq_HX},
\begin{equation}\label{eq_Hamilton_Eq}
\dot{ N} = E_{J} \sin(\phi) \qquad
\dot{\phi} = -2E_{C} \left( N + N_\text{env} \right)\ . 
\end{equation}
Eliminating the equation for the number of Cooper pairs $ N$, we get
\begin{equation}
\begin{split}
&\ddot{\phi} = -2E_{C}E_{J} \sin(\phi) - 2E_{C} \dot{N}_\text{env}\ ,
\\
&\dot{N}_\text{env} = i [H_\text{env},N_\text{env}]_{-}\ ,
\end{split}
\end{equation}
where the phase commutes with the environment operators:
$[\dot{\phi},N_\text{env}]_{-}=0$.
All operators are given in the Heisenberg representation, namely 
\begin{equation}
A\equiv A(t) = e^{iHt}A(0)e^{-iHt},\qquad H(t)=H(0)\ .
\end{equation}
Separating the coupling, $H=H_{0}+\ldots$, and switching to the interaction representation $A_{i}(t) = e^{iH_{0}t}A(0)e^{-iH_{0}t}$ (which makes $\dot{\phi}$ commute with all environment operators) we get, expanding the time-ordered propagator 
\begin{equation}
    \ddot{\phi} = -2E_{C}E_{J} \sin(\phi) - E_{C} J + E_{C} \int_{t_0}^{t} dt' Y(t-t')\dot{\phi}(t')\ ,
\end{equation}
with
\begin{equation}
\begin{split}
&J = 2i[H_\text{env}(t),N_\text{env}(t)]_{-}\ ,\\
&Y(t-t') = 2[  N_\text{env}(t') ,\,[H_\text{env}(t),N_\text{env}(t)]_{-}]_{-} \ ,
\end{split}
\label{eq_JY_def}
\end{equation}
where all time-dependency of operators are given in terms of the decoupled system with Hamiltonian $H_\text{env}'=H_\text{env}- E_{C} N_\text{env}^{2}$.

Averaging out over the external degrees of freedom, we get
\begin{align}
J =&\, 2i Z^{-1}\text{Tr} \Bigl\{e^{-\beta H_\text{env}'} [H_\text{env},N_\text{env}]_{-} \Bigr\}\ ,\\\nonumber
Y(t)=&\, 2Z^{-1}\text{Tr} \Bigl\{e^{-\beta H_\text{env}'} \times \\
&[ N_\text{env},\,e^{-i H_\text{env}'t}[H_\text{env},N_\text{env}]_{-}e^{i H_\text{env}'t}]_{-} \Bigr\}\ ,
\label{eq_Y_gen_def}
\end{align}
with the partition function $Z = \text{Tr} e^{-\beta H_\text{env}'}$.
Assuming fast internal dynamics of the bath, we can simplify the dissipation term to
\begin{equation}
 \int_{t_0}^{t} dt' Y(t-t')\dot{\phi}(t') \approx - \dot{\phi}(t)\int_{0}^{t-t_{0}} dt' Y(t') \ .
\end{equation}
Let us apply these general formulae for the particular models.

\subsection{Caldeira-Leggett model}

%
Representing the bath degrees of freedom in the Caldeira-Leggett model given in the Eq.~\eqref{eq_HB} by means of the bosonic ladder operators $a$, $a^{\dagger}$, we get the expression
\begin{equation}
N_\text{env} = \sum_{k}\lambda^z_{k}(a_{k}+a_{k}^{\dagger})\ ,\quad  H_\text{env} = \sum_{k}\omega_{k}a_{k}^{\dagger}a_{k}\ ,
\end{equation}
where $\omega_{k}=1/{\sqrt{L_{k}C_{k}}}$, and $\lambda^z_{k}=\lambda_{k}\sqrt{\frac{C_{k}\omega_{k}}{2}}=\lambda_{k}\sqrt[4]{\frac{C_{k}}{4L_{k}}}$ is not the same as the coupling in the initial Eq.~\eqref{eq_CL}.
The commutators given in the Eq.~\eqref{eq_JY_def} in the interaction representation are
\begin{align*}
[H_\text{env},N_\text{env}](t) \!=\! \sum_{k}\!\lambda^z_{k}\omega_{k} (-a_{k}e^{-i\omega_{k}t} \!+\! a_{k}^{\dagger}e^{i\omega_{k}t})\ , 
\\
[ N_\text{env}, [H_\text{env},N_\text{env}](t) ] \!=\! \sum_{k}\!|\lambda^z_{k}|^{2}\omega_{k} (e^{-i\omega_{k}t} \!+\! e^{i\omega_{k}t}) \!=\! Y(t)\ .
\end{align*}
Taking the trace over the ladder operators and assuming that the system is already become stationary, i.e.\ $t_{0}\to-\infty$, we get
\begin{align}
R^{-1} =& \int_{0}^{\infty}Y(\tau)d\tau = \pi\sum_{k}|\lambda^z_{k}|^{2}\omega_{k}\delta(\omega_{k}) \\=& \frac{\pi}{2} \sum_{k} C_{k}|\lambda_{k}|^{2} \omega_{k}^{2}\delta(\omega_{k}) = \frac{\pi}{2}\sum_{k} \frac{|\lambda_{k}|^{2}}{L_{k}}\delta(\omega_{k})\ ,
\end{align}
as in Eq.~\eqref{eq_R_CL} in the main text.
%

\subsection{Fermionic model}

Since both the Caldeira-Leggett and fermionic models can be cast into the same form, Eq.~\eqref{eq_HX}, the Hamilton equations are initially the same as in Eq.~\eqref{eq_Hamilton_Eq}.
To proceed, we use the expansion approach,
\begin{multline}
[N_\text{env}, H_\text{env}] = \sum_{kk'} \Biggl(\lambda^{\perp}_{kk'} c_{k'\downarrow}c_{k\uparrow} - \lambda^{\perp*}_{kk'} c_{k\uparrow}^{\dagger}c_{k'\downarrow}^{\dagger} 
 \\ 
 +\sum_{\sigma} (\epsilon_{k'}-\epsilon_{k})\lambda^{z}_{kk'}c_{k\sigma}^{\dagger}c_{k'\sigma}\Biggr)\ ,
\end{multline}
The trace in the Eq.~\eqref{eq_Y_gen_def} can be rewritten in the form, which simplifies the further calculation: $\text{Tr}\,\left\{ \int_{0}^{\infty}[N_\text{env}, H_\text{env}](t)e^{i\omega t}dt \, [N_\text{env}, \rho_\text{env}] \right\}$.
The components of the trace are
\begin{multline}
\int_{0}^{\infty}[N_\text{env}, H_\text{env}](t)e^{i\omega t}dt = 
 \sum_{kk'} \Biggl(
 \\
\sum_{\sigma}(\epsilon_{k'}-\epsilon_{k})\lambda_{kk'}^{z}c_{k\sigma}^{\dagger}c_{k'\sigma}\frac{1}{i(-\omega+\epsilon_{k}-\epsilon_{k'})+\delta}
 \\
 +
 \lambda^{\perp}_{kk'} c_{k'\downarrow}c_{k\uparrow}\frac{1}{-i(\omega + \epsilon_{k}+\epsilon_{k'})+\delta} 
 \\
 - \lambda^{\perp*}_{kk'} c_{k\uparrow}^{\dagger}c_{k'\downarrow}^{\dagger} \frac{1}{i(-\omega+\epsilon_{k}+\epsilon_{k'})+\delta}
\Biggr)\ .
\end{multline}
and commutator with density matrix can be calculated as 
\begin{multline}
[N_\text{env}, \rho_\text{env}]
=
Z^{-1} 
\sum_{kk'} \Biggl(
\\
\sum_{\sigma}(\epsilon_{k'}-\epsilon_{k})\lambda_{kk'}^{z}c_{k\sigma}^{\dagger}c_{k'\sigma}\frac{e^{\beta(\epsilon_{k'}-\epsilon_{k})}-1}{\epsilon_{k'}-\epsilon_{k}}
\\
+
\lambda^{\perp}_{kk'} c_{k'\downarrow}c_{k\uparrow}
\frac{e^{\beta(\epsilon_{k}+\epsilon_{k'})}-1}{\epsilon_{k}+\epsilon_{k'}} 
\\
+
\lambda^{\perp*}_{kk'} c_{k\uparrow}^{\dagger}c_{k'\downarrow}^{\dagger}
\frac{e^{-\beta(\epsilon_{k}+\epsilon_{k'})}-1}{\epsilon_{k}+\epsilon_{k'}} 
\Biggr)
e^{-\beta \sum_{k}c_{k\sigma}^{\dagger}c_{k\sigma}}\ .
\end{multline}
Splitting into the real and imaginary part of the trace we get (here we ignore the spin dependence)
\begin{align*}
\text{Re} Y(\omega) &= \pi\sum_{kk'} \Biggl( 
\\
&|\lambda^{\perp}_{kk'}|^{2} (n_{k'}+n_{k}-1) 
\sum_{\pm} \delta(\omega\pm(\epsilon_{k}+\epsilon_{k'})) \frac{1}{\epsilon_{k}+\epsilon_{k'}}
\\
&+ \lambda_{kk'}^{z}\lambda_{k'k}^{z} (n_{k'}-n_{k}) \sum_{\pm} (\mp\omega) \delta(\omega\pm (\epsilon_{k}-\epsilon_{k'}))
\Biggr)\ ,
\\
\text{Im} Y(\omega) &= 2\pi\omega\sum_{kk'} \Biggl( 
\\
&|\lambda^{\perp}_{kk'}|^{2} (n_{k'}+n_{k}-1) 
\sum_{\pm} \frac{1}{\omega^{2}-(\epsilon_{k}+\epsilon_{k'})^{2}} \frac{1}{\epsilon_{k}+\epsilon_{k'}}
\\
&+ \lambda_{kk'}^{z}\lambda_{k'k}^{z} (n_{k'}-n_{k}) \sum_{\pm}  \frac{\epsilon_{k'}-\epsilon_{k}}{\omega^{2}- (\epsilon_{k}-\epsilon_{k'})^{2}} 
\Biggr)\ .
\end{align*}
Taking the adiabatic limit of $\omega\to 0$ and using the designation $\lambda^{z/\perp}_{kk'} = \lambda^{z/\perp}(\omega_{k}-\omega_{k'})$ we get
\begin{equation}
R^{-1}=Y(0) = \pi |\lambda^{\perp}(0)|^{2} \nu^{2} + \pi \nu^{2} \lim_{\omega\to0}(\lambda^{z}(\omega)\omega)^{2}\ ,
\end{equation}
the result we cite in Eq.~\eqref{eq_R_SN} of the main text.

\section{A question of charge counting}\label{apx:chargecounting}

In the main text, we observed that in the transmon regime, the task of identifying dissipative phase transitions is equivalent to a full-counting statistics problem, as the formula  
\begin{equation}\label{eq_ES_ren_F}
\left<e^{2\pi i (N_\text{loc}+N_\text{tot})}\right>=\left<e^{2\pi iN_\text{loc}}\right>.
\end{equation}
is structurally the same as the moment generating function $m(\chi)=\text{tr}[e^{i\chi N_\text{loc}}]$ with the counting field $\chi$ (with the difference, that we here only trace over a finite energy range). Consequently, the Ohmic nature of the bath directly relates to a fundamental full-counting statistics result. For illustration purposes, let us take the normal metal to be 1D, and the SN interface to be a parallel plate capacitor of length $l$. Then the charge right underneath the superconductor is the relevant one which needs to enter Eq.~\eqref{eq_ES_ren_F}. The computation thus reduces to the well-studied problem of electron counting in a finite interval $l$ on a line~\cite{Aristov_1998,Ivanov_2013,kashuba2023counting}.


These studies~\cite{Aristov_1998,Abanov2011,LeHur2012,Ivanov_2013,Ivanov_2016} of the $\langle e^{i\chi n}\rangle$, where $n$ is the charge on the interval $[0..l]$, showed that for a fuzzy (nonquantized) charge, the dependence on auxiliary field $\chi$ can be sufficiently described by the first two cumulants.
We stress that this is not the expansion in $\chi$, but is due to the infrared cut off on $l^{-1}$ (the interval length) which are present only in these two cumulants.
Let us calculate for general $\chi$ for $N_\text{loc}$, namely $e^{i\chi N_\text{loc}}$.
The first cumulant is proportional to the large ratio of the ultraviolet (Fermi momentum) and infrared cutoffs, but contributes only in the form of the oscillatory dependence on $\chi$.
\begin{equation*}
\langle N_\text{loc}\rangle =  \sum_{pp'\sigma}\lambda_{p-p'} \wick{ \c1 c_{p\sigma}^{\dagger} \c1 c_{p'\sigma}} 
= 2 \lambda_{0} \sum_{p} n_{p} 
\end{equation*}
The second cumulant has a weaker, logarithmic dependence on this ratio
\begin{multline*}
 \langle N_\text{loc}^{2}\rangle - \langle N_\text{loc}\rangle^{2}
=\\= 
\sum_{p_{1}p_{2}p_{3}p_{4}\sigma\sigma'}\lambda_{p_{1}-p_{2}}\lambda_{p_{3}-p_{4}} \wick{ \c1 c_{p_{1}\sigma}^{\dagger} \c2 c_{p_{2}\sigma} \c2 c_{p_{3}\sigma'}^{\dagger} \c1 c_{p_{4}\sigma'}}
=\\= 2\sum_{pp'}\lambda_{p-p'}\lambda_{p'-p} n_{p}(1-n_{p}),
\end{multline*}
as is shown in the main text.
Thus, using the modified expression for the $N_\text{loc}$ from Eq.~\eqref{eq_NF} and setting $\chi$ to $2\pi$, we get that the renormalization of the absolute value of the $E_{S}$ in Eq.~\eqref{eq_ES_ren_F}, which will be
\begin{equation}
    E_S\rightarrow 
    \left(\frac{\Lambda}{\Lambda_{0}}\right)^{|\lambda^{z}_{0}|^{2}\times\text{const}}E_{S}\ ,
\end{equation}
and the constant power factor here hints that the bath is Ohmic.
Again, note that the terms with $\lambda^{\perp}$ from Eq.~\eqref{eq_HF} cannot contribute to the renormalization, as they merely form a proximity gap at low energies and thus may contribute to the $\Lambda$, but not to the exponent.
We showed above that for $\chi=2\pi$, the contribution to $N_\text{tot}$ vanishes, since the operator has the integer eigenvalues only.
Looking at the second cumulant, we find that it is zero for $N_\text{tot}$ substituting $\lambda_{p-p'}\to\delta_{pp'}$, what allows us to neglect this term also for general $\chi$.


To complete this discussion, we touch upon the matter of the interaction and disorder.
As Aristov demonstrated in his work~\cite{Aristov_1998} using a bosonization approach, the e-e interaction in 1D case indeed creates a multiplicative factor $K$, the Luttinger parameter, in front of the renormalization exponent $2K(\lambda^{z}_{0})^{2}$, though in higher dimensions, the interaction leads rather to the formation of the Fermi liquid, and, thus, does not affect the phase transition.
Disorder is even less relevant in this case, as it merely reconstructs the energy levels while leaving density of states, and thus the renormalization exponent untouched. We thus conclude that the normal metal resistivity hardly plays a role for the SB transition; instead the phase transition is dominantly controlled by the strength of the SN capacitance, which defines the magnitude of $\lambda^z_{0}$, see Fig.~\ref{fig_main}d.

\section{Bosonization of the Hamiltonian and comparison to other bosonic models}\label{apx:bosonization}

As already indicated in the introduction of the main text, contrary to our finding, Ref.~\cite{Schoen_1990} concluded the original Caldeira-Legget model, Eq.~\eqref{eq_CL}, to be suitable to describe dissipation through coupling with a normal metal. Since Ref.~\cite{Schoen_1990} applied a transformation to a bosonized representation of the normal metal degrees of freedom (via Hubbard-Stratonovich~\cite{Belitz_1994}), we will likewise represent the SN coupling in a bosonized way.

In fact, to set the stage, let us first consider a related bosonic model of a charge qubit electrically coupled to a Josephson junction array (the device measured in Ref.~\cite{kuzmin2023observation}). Assuming dominant capacitances to ground, the Hamiltonian here reads~\cite{Catelani_2011},
\begin{equation}
\begin{split}
    H=E_C N^2-E_J \cos(\phi)-e_J \cos(\phi_1-\phi)\\+\sum_j e_C N_j^2-e_J \cos(\phi_{j+1}-\phi_j)\ ,
\end{split}
\end{equation}
where $j$ indexes the islands of the array.
Notice that this model respects local charge quantization on each island ($\phi$ as well as all $\phi_j$ are compact). Here the electric contact manifests itself in the coupling $\sim\cos(\phi_1-\phi)$ exchanging Cooper pairs between the charge qubit and the first island, $j=1$. In analogy with the previous discussion, we can again transform this coupling into an effective capacitive interaction, Eq.~\eqref{eq_HX}, here, with the unitary $U^{(\text{JJA})}=e^{i\phi\sum_j N_j}$, yielding here an effective environment charge $N_\text{env}^\text{({JJA})}=\sum_j N_j$. This charge is quantized in the exact same manner as $N_\text{tot}$ in Eq.~\eqref{eq_NF}, such this coupling is likewise irrelevant for the SB transition in the regime $E_J>E_C$ (at least according to adiabatic renormalization). In order for the model to host an insulating phase, we likewise need a true electrostatic coupling, e.g., of the form $N_\text{env}^\text{(JJA)}=\sum_j \alpha_j N_j$. Consider now the fact that neighboring islands likely have a very small phase difference. One may thus be tempted to simply approximate all the cosines $\sim e_J$ quadratically. This leads us right to the Caldeira-Leggett model, Eq.~\eqref{eq_CL} (with an appropriate diagonalization of the array modes, that is). Crucially, we now get a finite renormalization (at least on the same level of adiabatic renormalization) due to the electric contact, which was absent before the approximation. This represents a marked difference due to a seemingly innocuous approximation step.

We now argue that the same subtle issue is in play when bosonizing the SN model (by means of the standard bosonization technique~\cite{Haldane_1981}), and attempting to map onto the CL model. To this end we first write the SN Hamiltonian in terms of continuous fermionic field operators,
\begin{equation}
\begin{split}
    H_{SN}=\int_0^L dx \Bigg[\sum_\sigma \Psi_\sigma^\dagger(x)\left(\frac{-\partial_x^2}{2m}\right)\Psi_\sigma(x)\\+\Delta(x)\left[e^{i\phi}\Psi_{\downarrow}\left(x\right)\Psi_{\uparrow}\left(x\right)+\text{h.c.}\right]\Bigg]\ ,
\end{split}
\end{equation}
where $L$ is the system size of the total heterostructure, and $\Psi_\sigma(x)$ is the standard fermionic field operator annihilating an electron with spin $\sigma$ at position $x$, satisfying the anti-commutation relation $\{\Psi_\sigma(x),\Psi_{\sigma^\prime}^\dagger (y)\}=\delta_{\sigma\sigma^\prime}\delta(x-y)$. The above Hamiltonian is similar to Eq.~\eqref{eq_H_SN_app}, except that here, we are specifically in 1D, and the SN interface is assumed to be ideal (no impurities at the interface). Assuming the interface to be located at $x=l$, the pairing is $\Delta(x)=\Delta$ for $0<x<l$ and $\Delta(x)=0$ for $l<x<L$.

Next, we cast the fermionic field operators into standard bosonized fields~\cite{Haldane_1981},
\begin{equation}
    \Psi_{\sigma}\left(x\right)=\sum_{\pm}\frac{1}{\sqrt{2\pi L}}e^{i\left[\varphi_{\sigma}\left(x\right)\pm\theta_{\sigma}\left(x\right)\right]}e^{\pm ik_{F}x}\ ,
\end{equation}
where $[n_\sigma(x),\varphi_{\sigma^\prime}(y)]=-i\delta_{\sigma\sigma^\prime}\delta(x-y)$ and $n_\sigma(x)=\partial_x \theta_\sigma$.
This yields the Hamiltonian
\begin{equation}
\begin{split}
 H_{SN}=\int_0^L dx\Bigg[ \frac{v_F}{2\pi}\sum_{\xi=n,s}\left([\partial_x \theta_\xi]^2+[\partial_x \varphi_\xi]^2\right)
    \\+\frac{2\Delta\left(x\right)}{\pi L}\cos\left[\phi+2\varphi_{n}\left(x\right)\right]\cos\left[\theta_{s}\left(x\right)\right]\Bigg]
\end{split}
\label{eq_H_SN_boson_sf}
\end{equation}
Here we have both a charge and spin boson field, $\theta_n=\theta_\uparrow+\theta_\downarrow$ and $\theta_s=\theta_\uparrow-\theta_\downarrow$ (and in analogy for $\varphi$, but with factor $\frac12$ to preserve the commutation relation).

For finite $\Delta$, we thus have in general spin dynamics on top of charge dynamics. This dynamics can be integrated out and  qualitatively the $\cos\left[\theta_{s}\left(x\right)\right]$ and be omitted. Note that the superconducting phase $\phi$ is compact on this level. 
Local charge quantization of the plasmon field is, however, a more involved subject~\cite{Haldane_1981,Rajaraman_1982,Aristov_1998,Pham_2000,Gutman2010,Ivanov_2013,Riwar_2021,kashuba2023counting}. In Eq.~\eqref{eq_H_SN_boson_sf}, we could in principle again transform the phase $\phi$ away by means of the unitary $e^{i\phi\int_0^L dx n(x)/2}$, leading yet again to a total charge shift $\sim \int_0^L dx n(x)=\theta(L)-\theta(0)$ which would appear in the capacitive coupling. The quantization of this total charge is not automatically guaranteed. But it can be ensured by applying boundary constraints on the boson field $\theta$ at $x=0,L$~\cite{Haldane_1981,Pham_2000}.  


In the large $\Delta$ limit, the charge field $\varphi_n(x)$ gets pinned to $\phi/2$ in the region with nonzero pairing ($0<x<l$), whereas the spin field gets pinned to the value $0$. 
Thus, we are left with a spin component which does not couple to the charge qubit, and the charge component of the boson field in the normal metal as a remaining dynamical degree of freedom. At the SN interface, $x=l$, the boson field must now satisfy the boundary constraint $\varphi(l)=\phi/2$.
This is the crucial step where charge quantization can get broken both in the superconductor as well as the normal metal: without a by-hand addition of a new, updated boundary condition for the boson field at $x=l$ and $x=L$, the total charge inside the normal metal is no longer quantized, since there is no natural constraint on the bosonic field $\theta$ at the other end of the normal metal. This can be fixed by hand imposing that the value $\theta(L)-\theta(l)$ has to be integer. In some sense it is equivalent to the quadratic expansion of the cosine in the JJ array transmission line example above: in the noninteracting boson limit (either $\cos(\phi)\approx 1-\phi^2/2$ or standard bosonization of fermionic fields by eliminating the lower bound of the Fermi sea~\cite{Haldane_1981}) one loses local charge quantization as the charge gets ``fuzzied out'' between superconductor and normal metal -- unless additional steps are taken (here in the form of boundary conditions) to requantize local charge. While these steps are for most purposes irrelevant, we show in the main text, that the eigenvalue properties of various charge operators play a crucial role for the RG procedure as formulated in Eq.~\eqref{eq_ES_coupling}.

\section{Josephson junctions array}\label{apx:jjarray}

Let us introduce a Josephson junction array model, which seems to be equivalent to the CL model in the limit of large Josephson energies:
\begin{multline}
L = - \frac{1}{4E_{C}} \dot{\phi}_{0}^{2} + E_{L} \phi_{0}^{2} - E_{J}\cos (\phi_{0} - \varphi) 
+ \\ + \sum_{m=1}^{N} \left(-\frac{1}{4e_{CN}} \dot{\phi}_{m}^{2} - \frac12 e_{JN}(\phi_{m} - \phi_{m-1})^{2}\right)
\end{multline}
where Matsubara imaginary time is used $\phi(\tau) = \sum_{\omega_{n}}e^{i\omega_{n}\tau}\phi(\omega_{n})$, with frequencies $\omega_{n}=2\pi T n$ and the sum is defined as $\sum_{\omega_{n}}\equiv T \sum_{n=-\infty}^{\infty}$.
Here we left the non-zero phase of the ground to use it in the calculations below.
If the mapping onto the CL model is correct, in case of the ohmic bath, the action of our model (after integrating all bath degrees of freedom) is
\begin{align}
S &= S_{0} + \int_{0}^{\beta}d\tau\,E_{J}\cos (\phi - \varphi),
\\
S_{0} &= \sum_{\omega_{n}}g^{-1}(\omega_{n}) \phi(\omega_{n})\phi(-\omega_{n})
\\
g^{-1}(\omega_{n}) &= \left(\frac{\omega_{n}^{2}}{4E_{C}} + E_{L} + \frac{\eta}{4\pi} |\omega_{n}| \right)
\end{align}
It time domain the bath contribution can be described by a term
\begin{equation}
\iint d\tau d\tau' K(\tau-\tau')\phi(\tau)\phi(\tau'), \qquad K(\tau) = \frac{\eta}{4\pi^{2} \tau^{2}}
\end{equation}
The integral of the action an be calculated %
using Wilson's approach, namely splitting the fast and the slow variables and integrating over the fast ones:
\begin{equation*}
e^{-S_{\text{eff}}} = e^{-S_{0,s}}Z_{0,f}^{-1}\int D \phi_{f}e^{-S_{0,f}}
e^{-\int d\tau\,E_{J}\cos (\phi(\tau) - \varphi(\tau)) }
\end{equation*}
The averaging of the exponent can be approximately calculated by expanding over the cumulants, i.e.\ $\langle e^{-\delta S_{f}} \rangle \approx  e^{-\langle\delta S_{f}\rangle + (\langle\delta S_{f}^{2}\rangle - \langle\delta S_{f}\rangle^{2} )/2} $.
The average is (here we incorporate the ground phase into the slow variables: $\phi_{s}\to\phi_{s}-\varphi$)
\begin{multline}
\langle\delta S_{f}\rangle = \left<\int d\tau\,E_{J}\cos (\phi(\tau) - \varphi(\tau))  \right> = \\= 
\frac12 E_{J} \int d\tau  \left< e^{-S_{0,f}}e^{i(\phi_{f}(\tau) + \phi_{s}(\tau))}\right>+ h.c.
= \\ = 
\frac12 E_{J} \int d\tau e^{i \phi_{s}(\tau) }  \left< e^{-S_{0,f}}e^{i\phi_{f}(\tau)}\right>+ h.c.
= \\ = 
 \frac12 E_{J}  e^{ - \sum_{\omega_{n}> 0}g(\omega_{n})/2} \int d\tau e^{i \phi_{s}(\tau) } +h.c.
\approx \\\approx
E_{J}  e^{ - \sum_{\omega_{n}> 0}g(\omega_{n})/2} \int d\tau \cos \phi_{s}(\tau) 
\end{multline}
Calculating the integral we get
\begin{equation}
e^{-S_{\text{eff}}} \approx e^{-S_{0,s}-  E_{J} b^{-1/\eta} \int d\tau \cos \phi_{s}(\tau)  }
\end{equation}
where $b = \Lambda_{0}/\Lambda$, and ``fast'' modes live inside $\Lambda < |\omega_{n}|<\Lambda_{0}$.
If we push the $\Lambda \to 0$ we get
$e^{-S_{\text{eff}}} = e^{-E_{J,\text{eff}}\int d\tau \cos \varphi(\tau)}$,
where 
$E_{J,\text{eff}} \sim E_{J} \left(\frac{\eta^{2} E_{C}}{E_{L}}\right)^{-1/\eta}$.
This is valid only for $\eta^{2} E_{C}\gg E_{L}$
The limit $\eta\to 0$ results in $E_{J,\text{eff}} \sim E_{J} e^{-\frac14\sqrt{\frac{E_{C}}{E_{L}}}}$, what pretty much means that small $\eta$ still implies $\eta \gg \sqrt{E_{L}/E_{C}}$.
This result constitutes the Schmid-Bulgadaev transition.
Further we will need also the second moment an the cumulant, which are
\begin{widetext}
\begin{multline*}
\langle\delta S_{f}^{2}\rangle = E_{J}^{2}\int d\tau d\tau'\left<\cos (\phi(\tau) - \varphi(\tau))\cos (\phi(\tau') - \varphi(\tau')) \right>
=\\=
\frac14E_{J}^{2}\int d\tau d\tau'\,\Bigl(
e^{i(\phi_{s}(\tau) + \phi_{s}(\tau')} e^{ - \sum_{\omega_{n}> 0}|1+e^{i\omega_{n}(\tau-\tau')}|^{2}g(\omega_{n})/2}
+
e^{i(\phi_{s}(\tau) - \phi_{s}(\tau'))} e^{ - \sum_{\omega_{n}> 0}|1-e^{i\omega_{n}(\tau-\tau')}|^{2}g(\omega_{n})/2} + h.c. \Bigr)
=\\=
\frac12E_{J}^{2}\int d\tau d\tau'\,\Bigl(
e^{ - 2\sum_{\omega_{n}> 0}\cos^{2}(\omega_{n}(\tau-\tau')/2)g(\omega_{n})} \cos(\phi_{s}(\tau) + \phi_{s}(\tau') )
+
e^{ - 2\sum_{\omega_{n}> 0}\sin^{2}(\omega_{n}(\tau-\tau')/2)g(\omega_{n})} \cos(\phi_{s}(\tau) - \phi_{s}(\tau'))
\end{multline*}
\begin{multline*}
\langle\delta S_{f}^{2}\rangle - \langle\delta S_{f}\rangle^{2} = 
\frac12 E_{J}^{2}e^{ - \sum_{\omega_{n}> 0}g(\omega_{n})} \int d\tau d\tau'\times\\\times\left(
[e^{ - \sum_{\omega_{n}> 0}\cos(\omega_{n}(\tau-\tau'))g(\omega_{n})} -1]\cos(\phi_{s}(\tau) + \phi_{s}(\tau'))
+
[e^{ \sum_{\omega_{n}> 0}\cos(\omega_{n}(\tau-\tau'))g(\omega_{n})} -1]\cos(\phi_{s}(\tau) - \phi_{s}(\tau'))\right)
\end{multline*}
\end{widetext}
Again, performing the whole integration ($\Lambda\to0$) and for $\eta^{2}E_{C}\gg E_{L}$ we get the correction to the effective action.
This result can be simplified by splitting into three timescales:
\begin{equation}
\delta S_{eff} = \frac14 E_{J}^{2} \int d\tau d\tau'\,K(\tau-\tau')\cos( \varphi(\tau)- \varphi(\tau'))
\end{equation}
where $K(\tau)=1$ for $|\tau|\ll 1/(\eta E_{C})$ and $K(\tau)=(\eta E_{C}|\tau-\tau'|)^{-2/\eta}$ for $1/(\eta E_{C})\ll|\tau|\ll \eta/E_{L}$.
For the large timescales $|\tau-\tau'|\gg \eta/E_{L}$ we get
\begin{equation}
\delta S_{eff} = \int d\tau d\tau'\,K(\tau-\tau')\sin \varphi(\tau)\sin\varphi(\tau')
\end{equation}
where
\begin{equation}
K(\tau) = \frac{\eta}{16\pi^{2}} \left(\frac{E_{J,\text{eff}}}{E_{L}}\right)^{2}|\tau-\tau'|^{-2}
\end{equation}
%
%
%


Let us now modify the system explicitly separating the first Josephson junction with energy $e_{J1}$, which will play the role of the bottleneck---the weak link between the bath and the quantum system.
The Lagrangian then looks like
\begin{multline}
L = - \frac{1}{4E_{C}} \dot{\phi}_{0}^{2} + E_{L} \phi_{0}^{2} - E_{J}\cos \phi_{0} - 
\\
- \frac{1}{4e_{C1}} \dot{\phi}_{1}^{2} + e_{L1} \phi_{1}^{2} - e_{J1}\cos (\phi_{1} - \phi_{0})
+\\+
 \sum_{m=2}^{N} \left(-\frac{1}{4e_{CN}} \dot{\phi}_{m}^{2} - \frac12 e_{JN}(\phi_{m} - \phi_{m-1})^{2}\right)
\end{multline}
Integrating all the bath variables, as we did above, we can reduce our systems action to
\begin{multline}
S=
\int d\tau\left(- \frac{1}{4E_{C}} \dot{\phi}^{2} + E_{L} \phi^{2} - E_{J}\cos \phi\right)
+\\+
\iint d\tau d\tau' K(\tau-\tau')\sin\phi(\tau)\sin\phi(\tau')
\end{multline}
where sines in the last term are the result of the phase compactness in the initial model.

\section{Bath contribution for the transmon regime}\label{apx:bath}

Here we consider the opposite---transmon case, when $E_{J}\ll E_{C}$.
To estimate the contribution to the bath we consider first the bare Lagrangian of the SC island:
\begin{equation}
L = - \frac{1}{4E_{C}} \dot{\phi}^{2}  - E_{J}\cos \phi.
\end{equation}
To differentiate the systems on closed manifold and CL model, we are interested in the dynamics of the field $\phi$, which leads the the transition from one minima ($\phi\approx 0$) to another ($\phi\approx 2\pi$), where the wavefunction is localized due to the large Josephson energy.
In this case we can exploit the instantonic approach ($t\to it$), which requires the a solution of the equation
\begin{equation}
E_{J} = \frac{1}{4E_{C}} \dot{\phi}^{2}  + E_{J}\cos \phi,
\end{equation}
that describes the transition on the infinite time interval $t \in (-\infty..\infty)$:
\begin{equation}
\phi = 4\arctan e^{\sqrt{2E_{C}E_{J}}t}.
\end{equation}
To compare the CL bath contribution and the corresponding term in the Josephson junction array model we simply substitute the expression for $\phi$ in the corresponding formulae, remembering that the kernel $K$ has more refined structure, namely
\begin{equation}
K(\tau) = \frac{\eta}{4\pi^{2}}\frac{\tau^{2}-\delta^{2}}{(\tau^{2}+\delta^{2})^{2}},
\quad\text{with $\delta\to0$.}
\end{equation}
This gives us an estimation of the bath contribution for the CL model
\begin{equation}
\iint d\tau d\tau' K(\tau-\tau')\phi(\tau)\phi(\tau') \approx -\frac{\eta}{16} \int_{0}^{\infty}\frac{dt}{t}
\end{equation}
where the divergent integral can be estimated as $b=\Lambda/\Lambda_{0}$.
For the Josephson junction array model we can perform the computation numerically obtaining
\begin{equation}
\iint d\tau d\tau' K(\tau-\tau')\sin\phi(\tau)\sin\phi(\tau') \approx -0.22 \eta.
\end{equation}
As we see, again, the JJ array model gives a significantly different prediction for the bath contribution.

\bibliography{paper1,paper2,references,biblio,bibliography}

\begin{thebibliography}{88}%
\makeatletter
\providecommand \@ifxundefined [1]{%
 \@ifx{#1\undefined}
}%
\providecommand \@ifnum [1]{%
 \ifnum #1\expandafter \@firstoftwo
 \else \expandafter \@secondoftwo
 \fi
}%
\providecommand \@ifx [1]{%
 \ifx #1\expandafter \@firstoftwo
 \else \expandafter \@secondoftwo
 \fi
}%
\providecommand \natexlab [1]{#1}%
\providecommand \enquote  [1]{``#1''}%
\providecommand \bibnamefont  [1]{#1}%
\providecommand \bibfnamefont [1]{#1}%
\providecommand \citenamefont [1]{#1}%
\providecommand \href@noop [0]{\@secondoftwo}%
\providecommand \href [0]{\begingroup \@sanitize@url \@href}%
\providecommand \@href[1]{\@@startlink{#1}\@@href}%
\providecommand \@@href[1]{\endgroup#1\@@endlink}%
\providecommand \@sanitize@url [0]{\catcode `\\12\catcode `\$12\catcode
  `\&12\catcode `\#12\catcode `\^12\catcode `\_12\catcode `\%12\relax}%
\providecommand \@@startlink[1]{}%
\providecommand \@@endlink[0]{}%
\providecommand \url  [0]{\begingroup\@sanitize@url \@url }%
\providecommand \@url [1]{\endgroup\@href {#1}{\urlprefix }}%
\providecommand \urlprefix  [0]{URL }%
\providecommand \Eprint [0]{\href }%
\providecommand \doibase [0]{https://doi.org/}%
\providecommand \selectlanguage [0]{\@gobble}%
\providecommand \bibinfo  [0]{\@secondoftwo}%
\providecommand \bibfield  [0]{\@secondoftwo}%
\providecommand \translation [1]{[#1]}%
\providecommand \BibitemOpen [0]{}%
\providecommand \bibitemStop [0]{}%
\providecommand \bibitemNoStop [0]{.\EOS\space}%
\providecommand \EOS [0]{\spacefactor3000\relax}%
\providecommand \BibitemShut  [1]{\csname bibitem#1\endcsname}%
\let\auto@bib@innerbib\@empty
\bibitem [{\citenamefont {Haldane}(1981)}]{Haldane_1981}%
  \BibitemOpen
  \bibfield  {author} {\bibinfo {author} {\bibfnamefont {F.~D.~M.}\
  \bibnamefont {Haldane}},\ }\bibfield  {title} {\bibinfo {title} {Luttinger
  liquid theory of one-dimensional quantum fluids. i. properties of the
  luttinger model and their extension to the general 1d interacting spinless
  fermi gas},\ }\href@noop {} {\bibfield  {journal} {\bibinfo  {journal}
  {Journal of Physics C: Solid State Physics}\ }\textbf {\bibinfo {volume}
  {14}},\ \bibinfo {pages} {2585} (\bibinfo {year} {1981})}\BibitemShut
  {NoStop}%
\bibitem [{\citenamefont {Pham}\ \emph {et~al.}(2000)\citenamefont {Pham},
  \citenamefont {Gabay},\ and\ \citenamefont {Lederer}}]{Pham_2000}%
  \BibitemOpen
  \bibfield  {author} {\bibinfo {author} {\bibfnamefont {K.-V.}\ \bibnamefont
  {Pham}}, \bibinfo {author} {\bibfnamefont {M.}~\bibnamefont {Gabay}},\ and\
  \bibinfo {author} {\bibfnamefont {P.}~\bibnamefont {Lederer}},\ }\bibfield
  {title} {\bibinfo {title} {Fractional excitations in the luttinger liquid},\
  }\href {https://doi.org/10.1103/PhysRevB.61.16397} {\bibfield  {journal}
  {\bibinfo  {journal} {Phys. Rev. B}\ }\textbf {\bibinfo {volume} {61}},\
  \bibinfo {pages} {16397} (\bibinfo {year} {2000})}\BibitemShut {NoStop}%
\bibitem [{\citenamefont {Altland}\ and\ \citenamefont
  {Simons}(2010)}]{Altland_Simons_book}%
  \BibitemOpen
  \bibfield  {author} {\bibinfo {author} {\bibfnamefont {A.}~\bibnamefont
  {Altland}}\ and\ \bibinfo {author} {\bibfnamefont {B.~D.}\ \bibnamefont
  {Simons}},\ }\href@noop {} {\emph {\bibinfo {title} {Condensed Matter Field
  Theory}}}\ (\bibinfo  {publisher} {Cambridge University Press},\ \bibinfo
  {address} {Cambridge, UK},\ \bibinfo {year} {2010})\BibitemShut {NoStop}%
\bibitem [{\citenamefont {Bruder}\ \emph {et~al.}(2005)\citenamefont {Bruder},
  \citenamefont {Fazio},\ and\ \citenamefont {Sch{\"o}n}}]{Bruder_2005}%
  \BibitemOpen
  \bibfield  {author} {\bibinfo {author} {\bibfnamefont {C.}~\bibnamefont
  {Bruder}}, \bibinfo {author} {\bibfnamefont {R.}~\bibnamefont {Fazio}},\ and\
  \bibinfo {author} {\bibfnamefont {G.}~\bibnamefont {Sch{\"o}n}},\ }\bibfield
  {title} {\bibinfo {title} {The bose-hubbard model: from josephson junction
  arrays to optical lattices},\ }\href
  {https://doi.org/10.1002/andp.200551709-1005} {\bibfield  {journal} {\bibinfo
   {journal} {Annalen der Physik}\ }\textbf {\bibinfo {volume} {517}},\
  \bibinfo {pages} {566} (\bibinfo {year} {2005})}\BibitemShut {NoStop}%
\bibitem [{\citenamefont {Koch}\ \emph {et~al.}(2009)\citenamefont {Koch},
  \citenamefont {Manucharyan}, \citenamefont {Devoret},\ and\ \citenamefont
  {Glazman}}]{Koch_2009}%
  \BibitemOpen
  \bibfield  {author} {\bibinfo {author} {\bibfnamefont {J.}~\bibnamefont
  {Koch}}, \bibinfo {author} {\bibfnamefont {V.}~\bibnamefont {Manucharyan}},
  \bibinfo {author} {\bibfnamefont {M.~H.}\ \bibnamefont {Devoret}},\ and\
  \bibinfo {author} {\bibfnamefont {L.~I.}\ \bibnamefont {Glazman}},\
  }\bibfield  {title} {\bibinfo {title} {Charging effects in the inductively
  shunted josephson junction},\ }\href
  {https://doi.org/10.1103/PhysRevLett.103.217004} {\bibfield  {journal}
  {\bibinfo  {journal} {Phys. Rev. Lett.}\ }\textbf {\bibinfo {volume} {103}},\
  \bibinfo {pages} {217004} (\bibinfo {year} {2009})}\BibitemShut {NoStop}%
\bibitem [{\citenamefont {Manucharyan}\ \emph {et~al.}(2009)\citenamefont
  {Manucharyan}, \citenamefont {Koch}, \citenamefont {Glazman},\ and\
  \citenamefont {Devoret}}]{Manucharyan_2009}%
  \BibitemOpen
  \bibfield  {author} {\bibinfo {author} {\bibfnamefont {V.~E.}\ \bibnamefont
  {Manucharyan}}, \bibinfo {author} {\bibfnamefont {J.}~\bibnamefont {Koch}},
  \bibinfo {author} {\bibfnamefont {L.~I.}\ \bibnamefont {Glazman}},\ and\
  \bibinfo {author} {\bibfnamefont {M.~H.}\ \bibnamefont {Devoret}},\
  }\bibfield  {title} {\bibinfo {title} {Fluxonium: Single cooper-pair circuit
  free of charge offsets},\ }\href {https://doi.org/10.1126/science.1175552}
  {\bibfield  {journal} {\bibinfo  {journal} {Science}\ }\textbf {\bibinfo
  {volume} {326}},\ \bibinfo {pages} {113} (\bibinfo {year}
  {2009})}\BibitemShut {NoStop}%
\bibitem [{\citenamefont {Catelani}\ \emph {et~al.}(2011)\citenamefont
  {Catelani}, \citenamefont {Schoelkopf}, \citenamefont {Devoret},\ and\
  \citenamefont {Glazman}}]{Catelani_2011}%
  \BibitemOpen
  \bibfield  {author} {\bibinfo {author} {\bibfnamefont {G.}~\bibnamefont
  {Catelani}}, \bibinfo {author} {\bibfnamefont {R.~J.}\ \bibnamefont
  {Schoelkopf}}, \bibinfo {author} {\bibfnamefont {M.~H.}\ \bibnamefont
  {Devoret}},\ and\ \bibinfo {author} {\bibfnamefont {L.~I.}\ \bibnamefont
  {Glazman}},\ }\bibfield  {title} {\bibinfo {title} {Relaxation and frequency
  shifts induced by quasiparticles in superconducting qubits},\ }\href
  {https://doi.org/10.1103/PhysRevB.84.064517} {\bibfield  {journal} {\bibinfo
  {journal} {Phys. Rev. B}\ }\textbf {\bibinfo {volume} {84}},\ \bibinfo
  {pages} {064517} (\bibinfo {year} {2011})}\BibitemShut {NoStop}%
\bibitem [{\citenamefont {Rajaraman}\ and\ \citenamefont
  {Bell}(1982)}]{Rajaraman_1982}%
  \BibitemOpen
  \bibfield  {author} {\bibinfo {author} {\bibfnamefont {R.}~\bibnamefont
  {Rajaraman}}\ and\ \bibinfo {author} {\bibfnamefont {J.}~\bibnamefont
  {Bell}},\ }\bibfield  {title} {\bibinfo {title} {On solitons with half
  integral charge},\ }\href
  {https://doi.org/https://doi.org/10.1016/0370-2693(82)90996-0} {\bibfield
  {journal} {\bibinfo  {journal} {Physics Letters B}\ }\textbf {\bibinfo
  {volume} {116}},\ \bibinfo {pages} {151} (\bibinfo {year}
  {1982})}\BibitemShut {NoStop}%
\bibitem [{\citenamefont {Riwar}(2021)}]{Riwar_2021}%
  \BibitemOpen
  \bibfield  {author} {\bibinfo {author} {\bibfnamefont {R.-P.}\ \bibnamefont
  {Riwar}},\ }\bibfield  {title} {\bibinfo {title} {Charge quantization and
  detector resolution},\ }\href {https://doi.org/10.21468/scipostphys.10.4.093}
  {\bibfield  {journal} {\bibinfo  {journal} {SciPost Physics}\ }\textbf
  {\bibinfo {volume} {10}},\ \bibinfo {pages} {093} (\bibinfo {year}
  {2021})}\BibitemShut {NoStop}%
\bibitem [{\citenamefont {Koliofoti}\ and\ \citenamefont
  {Riwar}(2023)}]{Koliofoti_2023}%
  \BibitemOpen
  \bibfield  {author} {\bibinfo {author} {\bibfnamefont {C.}~\bibnamefont
  {Koliofoti}}\ and\ \bibinfo {author} {\bibfnamefont {R.-P.}\ \bibnamefont
  {Riwar}},\ }\bibfield  {title} {\bibinfo {title} {Compact description of
  quantum phase slip junctions},\ }\href
  {https://doi.org/10.1038/s41534-023-00790-w} {\bibfield  {journal} {\bibinfo
  {journal} {npj Quantum Information}\ }\textbf {\bibinfo {volume} {9}},\
  \bibinfo {pages} {125} (\bibinfo {year} {2023})}\BibitemShut {NoStop}%
\bibitem [{\citenamefont {Aristov}(1998)}]{Aristov_1998}%
  \BibitemOpen
  \bibfield  {author} {\bibinfo {author} {\bibfnamefont {D.~N.}\ \bibnamefont
  {Aristov}},\ }\bibfield  {title} {\bibinfo {title} {Bosonization for a
  wigner-jordan-like transformation: Backscattering and umklapp processes on a
  fictitious lattice},\ }\href {https://doi.org/10.1103/PhysRevB.57.12825}
  {\bibfield  {journal} {\bibinfo  {journal} {Phys. Rev. B}\ }\textbf {\bibinfo
  {volume} {57}},\ \bibinfo {pages} {12825} (\bibinfo {year}
  {1998})}\BibitemShut {NoStop}%
\bibitem [{\citenamefont {Gutman}\ \emph {et~al.}(2010)\citenamefont {Gutman},
  \citenamefont {Gefen},\ and\ \citenamefont {Mirlin}}]{Gutman2010}%
  \BibitemOpen
  \bibfield  {author} {\bibinfo {author} {\bibfnamefont {D.~B.}\ \bibnamefont
  {Gutman}}, \bibinfo {author} {\bibfnamefont {Y.}~\bibnamefont {Gefen}},\ and\
  \bibinfo {author} {\bibfnamefont {A.~D.}\ \bibnamefont {Mirlin}},\ }\bibfield
   {title} {\bibinfo {title} {Full counting statistics of a luttinger liquid
  conductor},\ }\href {https://doi.org/10.1103/PhysRevLett.105.256802}
  {\bibfield  {journal} {\bibinfo  {journal} {Phys. Rev. Lett.}\ }\textbf
  {\bibinfo {volume} {105}},\ \bibinfo {pages} {256802} (\bibinfo {year}
  {2010})}\BibitemShut {NoStop}%
\bibitem [{\citenamefont {Ivanov}\ \emph {et~al.}(2013)\citenamefont {Ivanov},
  \citenamefont {Abanov},\ and\ \citenamefont {Cheianov}}]{Ivanov_2013}%
  \BibitemOpen
  \bibfield  {author} {\bibinfo {author} {\bibfnamefont {D.~A.}\ \bibnamefont
  {Ivanov}}, \bibinfo {author} {\bibfnamefont {A.~G.}\ \bibnamefont {Abanov}},\
  and\ \bibinfo {author} {\bibfnamefont {V.~V.}\ \bibnamefont {Cheianov}},\
  }\bibfield  {title} {\bibinfo {title} {Counting free fermions on a line: a
  fisher{\textendash}hartwig asymptotic expansion for the toeplitz determinant
  in the double-scaling limit},\ }\href
  {https://doi.org/10.1088/1751-8113/46/8/085003} {\bibfield  {journal}
  {\bibinfo  {journal} {Journal of Physics A: Mathematical and Theoretical}\
  }\textbf {\bibinfo {volume} {46}},\ \bibinfo {pages} {085003} (\bibinfo
  {year} {2013})}\BibitemShut {NoStop}%
\bibitem [{\citenamefont {Ivanov}\ and\ \citenamefont
  {Levkivskyi}(2016)}]{Ivanov_2016}%
  \BibitemOpen
  \bibfield  {author} {\bibinfo {author} {\bibfnamefont {D.~A.}\ \bibnamefont
  {Ivanov}}\ and\ \bibinfo {author} {\bibfnamefont {I.~P.}\ \bibnamefont
  {Levkivskyi}},\ }\bibfield  {title} {\bibinfo {title} {Fermionic full
  counting statistics with smooth boundaries: From discrete particles to
  bosonization},\ }\href {https://doi.org/10.1209/0295-5075/113/17009}
  {\bibfield  {journal} {\bibinfo  {journal} {Europhysics Letters}\ }\textbf
  {\bibinfo {volume} {113}},\ \bibinfo {pages} {17009} (\bibinfo {year}
  {2016})}\BibitemShut {NoStop}%
\bibitem [{\citenamefont {Kashuba}\ \emph {et~al.}(2023)\citenamefont
  {Kashuba}, \citenamefont {Schmidt}, \citenamefont {Hassler}, \citenamefont
  {Haller},\ and\ \citenamefont {Riwar}}]{kashuba2023counting}%
  \BibitemOpen
  \bibfield  {author} {\bibinfo {author} {\bibfnamefont {O.}~\bibnamefont
  {Kashuba}}, \bibinfo {author} {\bibfnamefont {T.~L.}\ \bibnamefont
  {Schmidt}}, \bibinfo {author} {\bibfnamefont {F.}~\bibnamefont {Hassler}},
  \bibinfo {author} {\bibfnamefont {A.}~\bibnamefont {Haller}},\ and\ \bibinfo
  {author} {\bibfnamefont {R.-P.}\ \bibnamefont {Riwar}},\ }\bibfield  {title}
  {\bibinfo {title} {Counting interacting electrons in one dimension},\ }\href
  {https://doi.org/10.1103/PhysRevB.108.235133} {\bibfield  {journal} {\bibinfo
   {journal} {Phys. Rev. B}\ }\textbf {\bibinfo {volume} {108}},\ \bibinfo
  {pages} {235133} (\bibinfo {year} {2023})}\BibitemShut {NoStop}%
\bibitem [{\citenamefont {Leggett}\ \emph {et~al.}(1987)\citenamefont
  {Leggett}, \citenamefont {Chakravarty}, \citenamefont {Dorsey}, \citenamefont
  {Fisher}, \citenamefont {Garg},\ and\ \citenamefont
  {Zwerger}}]{Leggett_1987}%
  \BibitemOpen
  \bibfield  {author} {\bibinfo {author} {\bibfnamefont {A.~J.}\ \bibnamefont
  {Leggett}}, \bibinfo {author} {\bibfnamefont {S.}~\bibnamefont
  {Chakravarty}}, \bibinfo {author} {\bibfnamefont {A.~T.}\ \bibnamefont
  {Dorsey}}, \bibinfo {author} {\bibfnamefont {M.~P.~A.}\ \bibnamefont
  {Fisher}}, \bibinfo {author} {\bibfnamefont {A.}~\bibnamefont {Garg}},\ and\
  \bibinfo {author} {\bibfnamefont {W.}~\bibnamefont {Zwerger}},\ }\bibfield
  {title} {\bibinfo {title} {Dynamics of the dissipative two-state system},\
  }\href {https://doi.org/10.1103/RevModPhys.59.1} {\bibfield  {journal}
  {\bibinfo  {journal} {Rev. Mod. Phys.}\ }\textbf {\bibinfo {volume} {59}},\
  \bibinfo {pages} {1} (\bibinfo {year} {1987})}\BibitemShut {NoStop}%
\bibitem [{\citenamefont {Giamarchi}\ and\ \citenamefont
  {Schulz}(1988)}]{Giamarchi_1988}%
  \BibitemOpen
  \bibfield  {author} {\bibinfo {author} {\bibfnamefont {T.}~\bibnamefont
  {Giamarchi}}\ and\ \bibinfo {author} {\bibfnamefont {H.~J.}\ \bibnamefont
  {Schulz}},\ }\bibfield  {title} {\bibinfo {title} {Anderson localization and
  interactions in one-dimensional metals},\ }\href
  {https://doi.org/https://doi.org/10.1103/PhysRevB.37.325} {\bibfield
  {journal} {\bibinfo  {journal} {Phys. Rev. B}\ }\textbf {\bibinfo {volume}
  {37}},\ \bibinfo {pages} {325} (\bibinfo {year} {1988})}\BibitemShut
  {NoStop}%
\bibitem [{Note1()}]{Note1}%
  \BibitemOpen
  \bibinfo {note} {To see explicit moment generating function structures
  pertinent to the RG flow, consider, e.g., Eqs.~(3.18)-(3.20) in Ref.~\cite
  {Leggett_1987}, or Appendix~A in Ref.~\cite {Giamarchi_1988}.}\BibitemShut
  {Stop}%
\bibitem [{\citenamefont {Devoret}()}]{Devoret_1997}%
  \BibitemOpen
  \bibfield  {author} {\bibinfo {author} {\bibfnamefont {M.~H.}\ \bibnamefont
  {Devoret}},\ }\href@noop {} {\bibinfo {title} {{Quantum Fluctuations, Les
  Houches, Session LXIII, edited by S. Reynaud, E. Giacobino, and J.
  Zinn-Justin (Elsevier Science, Amsterdam, 1997), pp. 351--386}}}\BibitemShut
  {NoStop}%
\bibitem [{\citenamefont {Burkard}\ \emph {et~al.}(2004)\citenamefont
  {Burkard}, \citenamefont {Koch},\ and\ \citenamefont
  {DiVincenzo}}]{Burkard_2004}%
  \BibitemOpen
  \bibfield  {author} {\bibinfo {author} {\bibfnamefont {G.}~\bibnamefont
  {Burkard}}, \bibinfo {author} {\bibfnamefont {R.~H.}\ \bibnamefont {Koch}},\
  and\ \bibinfo {author} {\bibfnamefont {D.~P.}\ \bibnamefont {DiVincenzo}},\
  }\bibfield  {title} {\bibinfo {title} {Multilevel quantum description of
  decoherence in superconducting qubits},\ }\href
  {https://doi.org/https://doi.org/10.1103/PhysRevB.69.064503} {\bibfield
  {journal} {\bibinfo  {journal} {Phys. Rev. B}\ }\textbf {\bibinfo {volume}
  {69}},\ \bibinfo {pages} {064503} (\bibinfo {year} {2004})}\BibitemShut
  {NoStop}%
\bibitem [{\citenamefont {Vool}\ and\ \citenamefont
  {Devoret}(2017)}]{Vool_2017}%
  \BibitemOpen
  \bibfield  {author} {\bibinfo {author} {\bibfnamefont {U.}~\bibnamefont
  {Vool}}\ and\ \bibinfo {author} {\bibfnamefont {M.}~\bibnamefont {Devoret}},\
  }\bibfield  {title} {\bibinfo {title} {Introduction to quantum
  electromagnetic circuits},\ }\href
  {https://doi.org/https://doi.org/10.1002/cta.2359} {\bibfield  {journal}
  {\bibinfo  {journal} {International Journal of Circuit Theory and
  Applications}\ }\textbf {\bibinfo {volume} {45}},\ \bibinfo {pages} {897}
  (\bibinfo {year} {2017})}\BibitemShut {NoStop}%
\bibitem [{\citenamefont {Riwar}\ and\ \citenamefont
  {DiVincenzo}(2022)}]{Riwar_2022}%
  \BibitemOpen
  \bibfield  {author} {\bibinfo {author} {\bibfnamefont {R.~P.}\ \bibnamefont
  {Riwar}}\ and\ \bibinfo {author} {\bibfnamefont {D.~P.}\ \bibnamefont
  {DiVincenzo}},\ }\bibfield  {title} {\bibinfo {title} {Circuit quantization
  with time-dependent magnetic fields for realistic geometries},\ }\href
  {https://doi.org/https://doi.org/10.1038/s41534-022-00539-x} {\bibfield
  {journal} {\bibinfo  {journal} {npj Quantum Information}\ }\textbf {\bibinfo
  {volume} {8}},\ \bibinfo {pages} {36} (\bibinfo {year} {2022})}\BibitemShut
  {NoStop}%
\bibitem [{\citenamefont {Peierls}(1979)}]{Peierls_book}%
  \BibitemOpen
  \bibfield  {author} {\bibinfo {author} {\bibfnamefont {R.}~\bibnamefont
  {Peierls}},\ }\href {http://www.jstor.org/stable/j.ctv1416425} {\emph
  {\bibinfo {title} {Surprises in Theoretical Physics}}}\ (\bibinfo
  {publisher} {Princeton University Press, Princeton},\ \bibinfo {year}
  {1979})\BibitemShut {NoStop}%
\bibitem [{\citenamefont {Likharev}\ and\ \citenamefont
  {Zorin}(1985)}]{Likharev_1985}%
  \BibitemOpen
  \bibfield  {author} {\bibinfo {author} {\bibfnamefont {K.~K.}\ \bibnamefont
  {Likharev}}\ and\ \bibinfo {author} {\bibfnamefont {A.~B.}\ \bibnamefont
  {Zorin}},\ }\bibfield  {title} {\bibinfo {title} {Theory of the bloch-wave
  oscillations in small josephson junctions},\ }\href
  {https://doi.org/10.1007/BF00683782} {\bibfield  {journal} {\bibinfo
  {journal} {Journal of Low Temperature Physics}\ }\textbf {\bibinfo {volume}
  {59}},\ \bibinfo {pages} {347} (\bibinfo {year} {1985})}\BibitemShut
  {NoStop}%
\bibitem [{\citenamefont {Cottet}(2002)}]{Cottet2002}%
  \BibitemOpen
  \bibfield  {author} {\bibinfo {author} {\bibfnamefont {A.}~\bibnamefont
  {Cottet}},\ }\emph {\bibinfo {title} {Implementation of a quantum bit in a
  superconducting circuit}},\ \href@noop {} {Ph.D. thesis},\ \bibinfo  {school}
  {Universit{\'e} Paris VI} (\bibinfo {year} {2002})\BibitemShut {NoStop}%
\bibitem [{\citenamefont {Koch}\ \emph {et~al.}(2007)\citenamefont {Koch},
  \citenamefont {Yu}, \citenamefont {Gambetta}, \citenamefont {Houck},
  \citenamefont {Schuster}, \citenamefont {Majer}, \citenamefont {Blais},
  \citenamefont {Devoret}, \citenamefont {Girvin},\ and\ \citenamefont
  {Schoelkopf}}]{Koch_2007}%
  \BibitemOpen
  \bibfield  {author} {\bibinfo {author} {\bibfnamefont {J.}~\bibnamefont
  {Koch}}, \bibinfo {author} {\bibfnamefont {T.~M.}\ \bibnamefont {Yu}},
  \bibinfo {author} {\bibfnamefont {J.}~\bibnamefont {Gambetta}}, \bibinfo
  {author} {\bibfnamefont {A.~A.}\ \bibnamefont {Houck}}, \bibinfo {author}
  {\bibfnamefont {D.~I.}\ \bibnamefont {Schuster}}, \bibinfo {author}
  {\bibfnamefont {J.}~\bibnamefont {Majer}}, \bibinfo {author} {\bibfnamefont
  {A.}~\bibnamefont {Blais}}, \bibinfo {author} {\bibfnamefont {M.~H.}\
  \bibnamefont {Devoret}}, \bibinfo {author} {\bibfnamefont {S.~M.}\
  \bibnamefont {Girvin}},\ and\ \bibinfo {author} {\bibfnamefont {R.~J.}\
  \bibnamefont {Schoelkopf}},\ }\bibfield  {title} {\bibinfo {title}
  {Charge-insensitive qubit design derived from the cooper pair box},\ }\href
  {https://doi.org/https://doi.org/10.1103/PhysRevA.76.042319} {\bibfield
  {journal} {\bibinfo  {journal} {Phys. Rev. A}\ }\textbf {\bibinfo {volume}
  {76}},\ \bibinfo {pages} {042319} (\bibinfo {year} {2007})}\BibitemShut
  {NoStop}%
\bibitem [{\citenamefont {Murani}\ \emph {et~al.}(2020)\citenamefont {Murani},
  \citenamefont {Bourlet}, \citenamefont {le~Sueur}, \citenamefont {Portier},
  \citenamefont {Altimiras}, \citenamefont {Esteve}, \citenamefont {Grabert},
  \citenamefont {Stockburger}, \citenamefont {Ankerhold},\ and\ \citenamefont
  {Joyez}}]{Murani_2020}%
  \BibitemOpen
  \bibfield  {author} {\bibinfo {author} {\bibfnamefont {A.}~\bibnamefont
  {Murani}}, \bibinfo {author} {\bibfnamefont {N.}~\bibnamefont {Bourlet}},
  \bibinfo {author} {\bibfnamefont {H.}~\bibnamefont {le~Sueur}}, \bibinfo
  {author} {\bibfnamefont {F.}~\bibnamefont {Portier}}, \bibinfo {author}
  {\bibfnamefont {C.}~\bibnamefont {Altimiras}}, \bibinfo {author}
  {\bibfnamefont {D.}~\bibnamefont {Esteve}}, \bibinfo {author} {\bibfnamefont
  {H.}~\bibnamefont {Grabert}}, \bibinfo {author} {\bibfnamefont
  {J.}~\bibnamefont {Stockburger}}, \bibinfo {author} {\bibfnamefont
  {J.}~\bibnamefont {Ankerhold}},\ and\ \bibinfo {author} {\bibfnamefont
  {P.}~\bibnamefont {Joyez}},\ }\bibfield  {title} {\bibinfo {title} {Absence
  of a dissipative quantum phase transition in josephson junctions},\ }\href
  {https://doi.org/10.1103/PhysRevX.10.021003} {\bibfield  {journal} {\bibinfo
  {journal} {Phys. Rev. X}\ }\textbf {\bibinfo {volume} {10}},\ \bibinfo
  {pages} {021003} (\bibinfo {year} {2020})}\BibitemShut {NoStop}%
\bibitem [{\citenamefont {Loss}\ and\ \citenamefont
  {Mullen}(1991)}]{Loss_1991}%
  \BibitemOpen
  \bibfield  {author} {\bibinfo {author} {\bibfnamefont {D.}~\bibnamefont
  {Loss}}\ and\ \bibinfo {author} {\bibfnamefont {K.}~\bibnamefont {Mullen}},\
  }\bibfield  {title} {\bibinfo {title} {Effect of dissipation on phase
  periodicity and the quantum dynamics of josephson junctions},\ }\href
  {https://doi.org/10.1103/PhysRevA.43.2129} {\bibfield  {journal} {\bibinfo
  {journal} {Phys. Rev. A}\ }\textbf {\bibinfo {volume} {43}},\ \bibinfo
  {pages} {2129} (\bibinfo {year} {1991})}\BibitemShut {NoStop}%
\bibitem [{\citenamefont {Mullen}\ \emph {et~al.}(1993)\citenamefont {Mullen},
  \citenamefont {Loss},\ and\ \citenamefont {Stoof}}]{Mullen_1993}%
  \BibitemOpen
  \bibfield  {author} {\bibinfo {author} {\bibfnamefont {K.}~\bibnamefont
  {Mullen}}, \bibinfo {author} {\bibfnamefont {D.}~\bibnamefont {Loss}},\ and\
  \bibinfo {author} {\bibfnamefont {H.~T.~C.}\ \bibnamefont {Stoof}},\
  }\bibfield  {title} {\bibinfo {title} {Resonant phenomena in compact and
  extended systems},\ }\href {https://doi.org/10.1103/PhysRevB.47.2689}
  {\bibfield  {journal} {\bibinfo  {journal} {Phys. Rev. B}\ }\textbf {\bibinfo
  {volume} {47}},\ \bibinfo {pages} {2689} (\bibinfo {year}
  {1993})}\BibitemShut {NoStop}%
\bibitem [{\citenamefont {Caldeira}\ and\ \citenamefont
  {Leggett}(1981)}]{Caldeira_1981}%
  \BibitemOpen
  \bibfield  {author} {\bibinfo {author} {\bibfnamefont {A.~O.}\ \bibnamefont
  {Caldeira}}\ and\ \bibinfo {author} {\bibfnamefont {A.~J.}\ \bibnamefont
  {Leggett}},\ }\bibfield  {title} {\bibinfo {title} {Influence of dissipation
  on quantum tunneling in macroscopic systems},\ }\href
  {https://doi.org/10.1103/PhysRevLett.46.211} {\bibfield  {journal} {\bibinfo
  {journal} {Phys. Rev. Lett.}\ }\textbf {\bibinfo {volume} {46}},\ \bibinfo
  {pages} {211} (\bibinfo {year} {1981})}\BibitemShut {NoStop}%
\bibitem [{Note2()}]{Note2}%
  \BibitemOpen
  \bibinfo {note} {In circuits, there exist the following main coupling types:
  charge-charge coupling due to capacitive interactions, current-current
  coupling, which is of inductive nature, and exchange of electrons across an
  interface. While capacitive interactions couple to the charge $N$, both
  inductive coupling and electron transfer usually couple to the phase $\phi $,
  such that one could in principle cast both of them under the umbrella term of
  ``inductive'' coupling. In our work, it will be of importance that the
  resistor allows for the transfer of charges, which is why we make explicitly
  this distinction.}\BibitemShut {Stop}%
\bibitem [{\citenamefont {Ingold}\ and\ \citenamefont
  {Nazarov}(1992)}]{Ingold:1992aa}%
  \BibitemOpen
  \bibfield  {author} {\bibinfo {author} {\bibfnamefont {G.-L.}\ \bibnamefont
  {Ingold}}\ and\ \bibinfo {author} {\bibfnamefont {Y.~V.}\ \bibnamefont
  {Nazarov}},\ }\bibinfo {title} {Charge tunneling rates in ultrasmall
  junctions},\ in\ \href {https://doi.org/10.1007/978-1-4757-2166-9_2} {\emph
  {\bibinfo {booktitle} {Single Charge Tunneling: Coulomb Blockade Phenomena In
  Nanostructures}}},\ \bibinfo {editor} {edited by\ \bibinfo {editor}
  {\bibfnamefont {H.}~\bibnamefont {Grabert}}\ and\ \bibinfo {editor}
  {\bibfnamefont {M.~H.}\ \bibnamefont {Devoret}}}\ (\bibinfo  {publisher}
  {Springer US},\ \bibinfo {address} {Boston, MA},\ \bibinfo {year} {1992})\
  pp.\ \bibinfo {pages} {21--107}\BibitemShut {NoStop}%
\bibitem [{\citenamefont {Schmid}(1983)}]{Schmid_1983}%
  \BibitemOpen
  \bibfield  {author} {\bibinfo {author} {\bibfnamefont {A.}~\bibnamefont
  {Schmid}},\ }\bibfield  {title} {\bibinfo {title} {Diffusion and localization
  in a dissipative quantum system},\ }\href
  {https://doi.org/10.1103/PhysRevLett.51.1506} {\bibfield  {journal} {\bibinfo
   {journal} {Phys. Rev. Lett.}\ }\textbf {\bibinfo {volume} {51}},\ \bibinfo
  {pages} {1506} (\bibinfo {year} {1983})}\BibitemShut {NoStop}%
\bibitem [{\citenamefont {Bulgadaev}(1984)}]{Bulgadaev_1984}%
  \BibitemOpen
  \bibfield  {author} {\bibinfo {author} {\bibfnamefont {S.}~\bibnamefont
  {Bulgadaev}},\ }\bibfield  {title} {\bibinfo {title} {Phase diagram of a
  dissipative quantum system},\ }\href
  {http://jetpletters.ru/ps/0/article_19477.shtml} {\bibfield  {journal}
  {\bibinfo  {journal} {Pis'ma Zh. Eksp. Teor. Fiz.}\ }\textbf {\bibinfo
  {volume} {39}},\ \bibinfo {pages} {264} (\bibinfo {year} {1984})},\ \bibinfo
  {note} {[JETP Lett., {\bf 39} (6), 315--319 (1984)]}\BibitemShut {NoStop}%
\bibitem [{\citenamefont {Aslangul}\ \emph {et~al.}(1985)\citenamefont
  {Aslangul}, \citenamefont {Pottier},\ and\ \citenamefont
  {Saint-James}}]{Aslangul_1985}%
  \BibitemOpen
  \bibfield  {author} {\bibinfo {author} {\bibfnamefont {C.}~\bibnamefont
  {Aslangul}}, \bibinfo {author} {\bibfnamefont {N.}~\bibnamefont {Pottier}},\
  and\ \bibinfo {author} {\bibfnamefont {D.}~\bibnamefont {Saint-James}},\
  }\bibfield  {title} {\bibinfo {title} {Quantum ohmic dissipation: Particle on
  a one-dimensional periodic lattice},\ }\href
  {https://doi.org/https://doi.org/10.1016/0375-9601(85)90570-5} {\bibfield
  {journal} {\bibinfo  {journal} {Physics Letters A}\ }\textbf {\bibinfo
  {volume} {111}},\ \bibinfo {pages} {175} (\bibinfo {year}
  {1985})}\BibitemShut {NoStop}%
\bibitem [{\citenamefont {Guinea}\ \emph {et~al.}(1985)\citenamefont {Guinea},
  \citenamefont {Hakim},\ and\ \citenamefont {Muramatsu}}]{Guinea_1985}%
  \BibitemOpen
  \bibfield  {author} {\bibinfo {author} {\bibfnamefont {F.}~\bibnamefont
  {Guinea}}, \bibinfo {author} {\bibfnamefont {V.}~\bibnamefont {Hakim}},\ and\
  \bibinfo {author} {\bibfnamefont {A.}~\bibnamefont {Muramatsu}},\ }\bibfield
  {title} {\bibinfo {title} {Diffusion and localization of a particle in a
  periodic potential coupled to a dissipative environment},\ }\href
  {https://doi.org/10.1103/PhysRevLett.54.263} {\bibfield  {journal} {\bibinfo
  {journal} {Phys. Rev. Lett.}\ }\textbf {\bibinfo {volume} {54}},\ \bibinfo
  {pages} {263} (\bibinfo {year} {1985})}\BibitemShut {NoStop}%
\bibitem [{\citenamefont {Sch\"on}\ and\ \citenamefont
  {Zaikin}(1990)}]{Schoen_1990}%
  \BibitemOpen
  \bibfield  {author} {\bibinfo {author} {\bibfnamefont {G.}~\bibnamefont
  {Sch\"on}}\ and\ \bibinfo {author} {\bibfnamefont {A.}~\bibnamefont
  {Zaikin}},\ }\bibfield  {title} {\bibinfo {title} {Quantum coherent effects,
  phase transitions, and the dissipative dynamics of ultra small tunnel
  junctions},\ }\href
  {https://doi.org/https://doi.org/10.1016/0370-1573(90)90156-V} {\bibfield
  {journal} {\bibinfo  {journal} {Physics Reports}\ }\textbf {\bibinfo {volume}
  {198}},\ \bibinfo {pages} {237 } (\bibinfo {year} {1990})}\BibitemShut
  {NoStop}%
\bibitem [{\citenamefont {Herrero}\ and\ \citenamefont
  {Zaikin}(2002)}]{Herrero_2002}%
  \BibitemOpen
  \bibfield  {author} {\bibinfo {author} {\bibfnamefont {C.~P.}\ \bibnamefont
  {Herrero}}\ and\ \bibinfo {author} {\bibfnamefont {A.~D.}\ \bibnamefont
  {Zaikin}},\ }\bibfield  {title} {\bibinfo {title} {Superconductor-insulator
  quantum phase transition in a single josephson junction},\ }\href
  {https://doi.org/10.1103/PhysRevB.65.104516} {\bibfield  {journal} {\bibinfo
  {journal} {Phys. Rev. B}\ }\textbf {\bibinfo {volume} {65}},\ \bibinfo
  {pages} {104516} (\bibinfo {year} {2002})}\BibitemShut {NoStop}%
\bibitem [{\citenamefont {Kimura}\ and\ \citenamefont
  {Kato}(2004)}]{Kimura_2004}%
  \BibitemOpen
  \bibfield  {author} {\bibinfo {author} {\bibfnamefont {N.}~\bibnamefont
  {Kimura}}\ and\ \bibinfo {author} {\bibfnamefont {T.}~\bibnamefont {Kato}},\
  }\bibfield  {title} {\bibinfo {title} {Temperature dependence of zero-bias
  resistances of a single resistance-shunted josephson junction},\ }\href
  {https://doi.org/10.1103/PhysRevB.69.012504} {\bibfield  {journal} {\bibinfo
  {journal} {Phys. Rev. B}\ }\textbf {\bibinfo {volume} {69}},\ \bibinfo
  {pages} {012504} (\bibinfo {year} {2004})}\BibitemShut {NoStop}%
\bibitem [{\citenamefont {Werner}\ and\ \citenamefont
  {Troyer}(2005)}]{Werner_2005}%
  \BibitemOpen
  \bibfield  {author} {\bibinfo {author} {\bibfnamefont {P.}~\bibnamefont
  {Werner}}\ and\ \bibinfo {author} {\bibfnamefont {M.}~\bibnamefont
  {Troyer}},\ }\bibfield  {title} {\bibinfo {title} {Efficient simulation of
  resistively shunted josephson junctions},\ }\href
  {https://doi.org/10.1103/PhysRevLett.95.060201} {\bibfield  {journal}
  {\bibinfo  {journal} {Phys. Rev. Lett.}\ }\textbf {\bibinfo {volume} {95}},\
  \bibinfo {pages} {060201} (\bibinfo {year} {2005})}\BibitemShut {NoStop}%
\bibitem [{\citenamefont {Lukyanov}\ and\ \citenamefont
  {Werner}(2007)}]{Lukyanov_2007}%
  \BibitemOpen
  \bibfield  {author} {\bibinfo {author} {\bibfnamefont {S.~L.}\ \bibnamefont
  {Lukyanov}}\ and\ \bibinfo {author} {\bibfnamefont {P.}~\bibnamefont
  {Werner}},\ }\bibfield  {title} {\bibinfo {title} {Resistively shunted
  josephson junctions: quantum field theory predictions versus monte carlo
  results},\ }\href {https://doi.org/10.1088/1742-5468/2007/06/P06002}
  {\bibfield  {journal} {\bibinfo  {journal} {Journal of Statistical Mechanics:
  Theory and Experiment}\ }\textbf {\bibinfo {volume} {2007}},\ \bibinfo
  {pages} {P06002} (\bibinfo {year} {2007})}\BibitemShut {NoStop}%
\bibitem [{\citenamefont {Hakonen}\ and\ \citenamefont
  {Sonin}(2021)}]{Hakonen_2021}%
  \BibitemOpen
  \bibfield  {author} {\bibinfo {author} {\bibfnamefont {P.~J.}\ \bibnamefont
  {Hakonen}}\ and\ \bibinfo {author} {\bibfnamefont {E.~B.}\ \bibnamefont
  {Sonin}},\ }\bibfield  {title} {\bibinfo {title} {Comment on ``absence of a
  dissipative quantum phase transition in josephson junctions''},\ }\href
  {https://doi.org/https://doi.org/10.1103/PhysRevX.11.018001} {\bibfield
  {journal} {\bibinfo  {journal} {Phys. Rev. X}\ }\textbf {\bibinfo {volume}
  {11}},\ \bibinfo {pages} {018001} (\bibinfo {year} {2021})}\BibitemShut
  {NoStop}%
\bibitem [{\citenamefont {Murani}\ \emph {et~al.}(2021)\citenamefont {Murani},
  \citenamefont {Bourlet}, \citenamefont {le~Sueur}, \citenamefont {Portier},
  \citenamefont {Altimiras}, \citenamefont {Esteve}, \citenamefont {Grabert},
  \citenamefont {Stockburger}, \citenamefont {Ankerhold},\ and\ \citenamefont
  {Joyez}}]{Murani_2021}%
  \BibitemOpen
  \bibfield  {author} {\bibinfo {author} {\bibfnamefont {A.}~\bibnamefont
  {Murani}}, \bibinfo {author} {\bibfnamefont {N.}~\bibnamefont {Bourlet}},
  \bibinfo {author} {\bibfnamefont {H.}~\bibnamefont {le~Sueur}}, \bibinfo
  {author} {\bibfnamefont {F.}~\bibnamefont {Portier}}, \bibinfo {author}
  {\bibfnamefont {C.}~\bibnamefont {Altimiras}}, \bibinfo {author}
  {\bibfnamefont {D.}~\bibnamefont {Esteve}}, \bibinfo {author} {\bibfnamefont
  {H.}~\bibnamefont {Grabert}}, \bibinfo {author} {\bibfnamefont
  {J.}~\bibnamefont {Stockburger}}, \bibinfo {author} {\bibfnamefont
  {J.}~\bibnamefont {Ankerhold}},\ and\ \bibinfo {author} {\bibfnamefont
  {P.}~\bibnamefont {Joyez}},\ }\bibfield  {title} {\bibinfo {title} {Reply to
  ``comment on `absence of a dissipative quantum phase transition in josephson
  junctions'''},\ }\href {https://doi.org/10.1103/PhysRevX.11.018002}
  {\bibfield  {journal} {\bibinfo  {journal} {Phys. Rev. X}\ }\textbf {\bibinfo
  {volume} {11}},\ \bibinfo {pages} {018002} (\bibinfo {year}
  {2021})}\BibitemShut {NoStop}%
\bibitem [{\citenamefont {Kuzmin}\ \emph {et~al.}()\citenamefont {Kuzmin},
  \citenamefont {Mehta}, \citenamefont {Grabon}, \citenamefont {Mencia},
  \citenamefont {Burshtein}, \citenamefont {Goldstein},\ and\ \citenamefont
  {Manucharyan}}]{kuzmin2023observation}%
  \BibitemOpen
  \bibfield  {author} {\bibinfo {author} {\bibfnamefont {R.}~\bibnamefont
  {Kuzmin}}, \bibinfo {author} {\bibfnamefont {N.}~\bibnamefont {Mehta}},
  \bibinfo {author} {\bibfnamefont {N.}~\bibnamefont {Grabon}}, \bibinfo
  {author} {\bibfnamefont {R.~A.}\ \bibnamefont {Mencia}}, \bibinfo {author}
  {\bibfnamefont {A.}~\bibnamefont {Burshtein}}, \bibinfo {author}
  {\bibfnamefont {M.}~\bibnamefont {Goldstein}},\ and\ \bibinfo {author}
  {\bibfnamefont {V.~E.}\ \bibnamefont {Manucharyan}},\ }\href@noop {}
  {}\Eprint {https://arxiv.org/abs/2304.05806} {arXiv:2304.05806} \BibitemShut
  {NoStop}%
\bibitem [{\citenamefont {Houzet}\ \emph {et~al.}(2024)\citenamefont {Houzet},
  \citenamefont {Yamamoto},\ and\ \citenamefont
  {Glazman}}]{houzet2023microwave}%
  \BibitemOpen
  \bibfield  {author} {\bibinfo {author} {\bibfnamefont {M.}~\bibnamefont
  {Houzet}}, \bibinfo {author} {\bibfnamefont {T.}~\bibnamefont {Yamamoto}},\
  and\ \bibinfo {author} {\bibfnamefont {L.~I.}\ \bibnamefont {Glazman}},\
  }\bibfield  {title} {\bibinfo {title} {Microwave spectroscopy of the schmid
  transition},\ }\href {https://doi.org/10.1103/PhysRevB.109.155431} {\bibfield
   {journal} {\bibinfo  {journal} {Phys. Rev. B}\ }\textbf {\bibinfo {volume}
  {109}},\ \bibinfo {pages} {155431} (\bibinfo {year} {2024})}\BibitemShut
  {NoStop}%
\bibitem [{\citenamefont {Burshtein}\ and\ \citenamefont
  {Goldstein}(2024)}]{burshtein2023inelastic}%
  \BibitemOpen
  \bibfield  {author} {\bibinfo {author} {\bibfnamefont {A.}~\bibnamefont
  {Burshtein}}\ and\ \bibinfo {author} {\bibfnamefont {M.}~\bibnamefont
  {Goldstein}},\ }\bibfield  {title} {\bibinfo {title} {Inelastic decay from
  integrability},\ }\href {https://doi.org/10.1103/PRXQuantum.5.020323}
  {\bibfield  {journal} {\bibinfo  {journal} {PRX Quantum}\ }\textbf {\bibinfo
  {volume} {5}},\ \bibinfo {pages} {020323} (\bibinfo {year}
  {2024})}\BibitemShut {NoStop}%
\bibitem [{\citenamefont {Masuki}\ \emph {et~al.}(2022)\citenamefont {Masuki},
  \citenamefont {Sudo}, \citenamefont {Oshikawa},\ and\ \citenamefont
  {Ashida}}]{Masuki_2022}%
  \BibitemOpen
  \bibfield  {author} {\bibinfo {author} {\bibfnamefont {K.}~\bibnamefont
  {Masuki}}, \bibinfo {author} {\bibfnamefont {H.}~\bibnamefont {Sudo}},
  \bibinfo {author} {\bibfnamefont {M.}~\bibnamefont {Oshikawa}},\ and\
  \bibinfo {author} {\bibfnamefont {Y.}~\bibnamefont {Ashida}},\ }\bibfield
  {title} {\bibinfo {title} {Absence versus presence of dissipative quantum
  phase transition in josephson junctions},\ }\href
  {https://doi.org/10.1103/PhysRevLett.129.087001} {\bibfield  {journal}
  {\bibinfo  {journal} {Phys. Rev. Lett.}\ }\textbf {\bibinfo {volume} {129}},\
  \bibinfo {pages} {087001} (\bibinfo {year} {2022})}\BibitemShut {NoStop}%
\bibitem [{\citenamefont {S\'epulcre}\ \emph {et~al.}(2023)\citenamefont
  {S\'epulcre}, \citenamefont {Florens},\ and\ \citenamefont
  {Snyman}}]{Sepulcre_2023}%
  \BibitemOpen
  \bibfield  {author} {\bibinfo {author} {\bibfnamefont {T.}~\bibnamefont
  {S\'epulcre}}, \bibinfo {author} {\bibfnamefont {S.}~\bibnamefont
  {Florens}},\ and\ \bibinfo {author} {\bibfnamefont {I.}~\bibnamefont
  {Snyman}},\ }\bibfield  {title} {\bibinfo {title} {Comment on ``absence
  versus presence of dissipative quantum phase transition in josephson
  junctions''},\ }\href {https://doi.org/10.1103/PhysRevLett.131.199701}
  {\bibfield  {journal} {\bibinfo  {journal} {Phys. Rev. Lett.}\ }\textbf
  {\bibinfo {volume} {131}},\ \bibinfo {pages} {199701} (\bibinfo {year}
  {2023})}\BibitemShut {NoStop}%
\bibitem [{\citenamefont {Masuki}\ \emph {et~al.}(2023)\citenamefont {Masuki},
  \citenamefont {Sudo}, \citenamefont {Oshikawa},\ and\ \citenamefont
  {Ashida}}]{Masuki_2023}%
  \BibitemOpen
  \bibfield  {author} {\bibinfo {author} {\bibfnamefont {K.}~\bibnamefont
  {Masuki}}, \bibinfo {author} {\bibfnamefont {H.}~\bibnamefont {Sudo}},
  \bibinfo {author} {\bibfnamefont {M.}~\bibnamefont {Oshikawa}},\ and\
  \bibinfo {author} {\bibfnamefont {Y.}~\bibnamefont {Ashida}},\ }\bibfield
  {title} {\bibinfo {title} {Masuki et al. reply:},\ }\href
  {https://doi.org/10.1103/PhysRevLett.131.199702} {\bibfield  {journal}
  {\bibinfo  {journal} {Phys. Rev. Lett.}\ }\textbf {\bibinfo {volume} {131}},\
  \bibinfo {pages} {199702} (\bibinfo {year} {2023})}\BibitemShut {NoStop}%
\bibitem [{\citenamefont {Giacomelli}\ and\ \citenamefont
  {Ciuti}(2024)}]{giacomelli2023emergent}%
  \BibitemOpen
  \bibfield  {author} {\bibinfo {author} {\bibfnamefont {L.}~\bibnamefont
  {Giacomelli}}\ and\ \bibinfo {author} {\bibfnamefont {C.}~\bibnamefont
  {Ciuti}},\ }\bibfield  {title} {\bibinfo {title} {Emergent quantum phase
  transition of a josephson junction coupled to a high-impedance multimode
  resonator},\ }\href {https://doi.org/10.1038/s41467-024-48558-w} {\bibfield
  {journal} {\bibinfo  {journal} {Nature Communications}\ }\textbf {\bibinfo
  {volume} {15}},\ \bibinfo {pages} {5455} (\bibinfo {year}
  {2024})}\BibitemShut {NoStop}%
\bibitem [{\citenamefont {Altimiras}\ \emph {et~al.}(2023)\citenamefont
  {Altimiras}, \citenamefont {Esteve}, \citenamefont {Girit}, \citenamefont
  {le~Sueur},\ and\ \citenamefont {Joyez}}]{altimiras2023absence}%
  \BibitemOpen
  \bibfield  {author} {\bibinfo {author} {\bibfnamefont {C.}~\bibnamefont
  {Altimiras}}, \bibinfo {author} {\bibfnamefont {D.}~\bibnamefont {Esteve}},
  \bibinfo {author} {\bibfnamefont {{\c C}.}~\bibnamefont {Girit}}, \bibinfo
  {author} {\bibfnamefont {H.}~\bibnamefont {le~Sueur}},\ and\ \bibinfo
  {author} {\bibfnamefont {P.}~\bibnamefont {Joyez}},\ }\href@noop {} {\bibinfo
  {title} {Absence of a dissipative quantum phase transition in josephson
  junctions: Theory}} (\bibinfo {year} {2023}),\ \Eprint
  {https://arxiv.org/abs/2312.14754} {arXiv:2312.14754 [cond-mat.supr-con]}
  \BibitemShut {NoStop}%
\bibitem [{\citenamefont {Subero}\ \emph {et~al.}(2023)\citenamefont {Subero},
  \citenamefont {Maillet}, \citenamefont {Golubev}, \citenamefont {Thomas},
  \citenamefont {Peltonen}, \citenamefont {Karimi}, \citenamefont
  {Mar{\'\i}n-Su{\'a}rez}, \citenamefont {Yeyati}, \citenamefont {S{\'a}nchez},
  \citenamefont {Park},\ and\ \citenamefont {Pekola}}]{subero2023bolometric}%
  \BibitemOpen
  \bibfield  {author} {\bibinfo {author} {\bibfnamefont {D.}~\bibnamefont
  {Subero}}, \bibinfo {author} {\bibfnamefont {O.}~\bibnamefont {Maillet}},
  \bibinfo {author} {\bibfnamefont {D.~S.}\ \bibnamefont {Golubev}}, \bibinfo
  {author} {\bibfnamefont {G.}~\bibnamefont {Thomas}}, \bibinfo {author}
  {\bibfnamefont {J.~T.}\ \bibnamefont {Peltonen}}, \bibinfo {author}
  {\bibfnamefont {B.}~\bibnamefont {Karimi}}, \bibinfo {author} {\bibfnamefont
  {M.}~\bibnamefont {Mar{\'\i}n-Su{\'a}rez}}, \bibinfo {author} {\bibfnamefont
  {A.~L.}\ \bibnamefont {Yeyati}}, \bibinfo {author} {\bibfnamefont
  {R.}~\bibnamefont {S{\'a}nchez}}, \bibinfo {author} {\bibfnamefont
  {S.}~\bibnamefont {Park}},\ and\ \bibinfo {author} {\bibfnamefont {J.~P.}\
  \bibnamefont {Pekola}},\ }\bibfield  {title} {\bibinfo {title} {Bolometric
  detection of josephson inductance in a highly resistive environment},\ }\href
  {https://doi.org/10.1038/s41467-023-43668-3} {\bibfield  {journal} {\bibinfo
  {journal} {Nature Communications}\ }\textbf {\bibinfo {volume} {14}},\
  \bibinfo {pages} {7924} (\bibinfo {year} {2023})}\BibitemShut {NoStop}%
\bibitem [{\citenamefont {Belitz}\ and\ \citenamefont
  {Kirkpatrick}(1994)}]{Belitz_1994}%
  \BibitemOpen
  \bibfield  {author} {\bibinfo {author} {\bibfnamefont {D.}~\bibnamefont
  {Belitz}}\ and\ \bibinfo {author} {\bibfnamefont {T.~R.}\ \bibnamefont
  {Kirkpatrick}},\ }\bibfield  {title} {\bibinfo {title} {The anderson-mott
  transition},\ }\href {https://doi.org/10.1103/RevModPhys.66.261} {\bibfield
  {journal} {\bibinfo  {journal} {Rev. Mod. Phys.}\ }\textbf {\bibinfo {volume}
  {66}},\ \bibinfo {pages} {261} (\bibinfo {year} {1994})}\BibitemShut
  {NoStop}%
\bibitem [{\citenamefont {Rollb\"uhler}\ and\ \citenamefont
  {Grabert}(2001)}]{Rollbuehler_2001}%
  \BibitemOpen
  \bibfield  {author} {\bibinfo {author} {\bibfnamefont {J.}~\bibnamefont
  {Rollb\"uhler}}\ and\ \bibinfo {author} {\bibfnamefont {H.}~\bibnamefont
  {Grabert}},\ }\bibfield  {title} {\bibinfo {title} {Coulomb blockade of
  tunneling between disordered conductors},\ }\href
  {https://doi.org/10.1103/PhysRevLett.87.126804} {\bibfield  {journal}
  {\bibinfo  {journal} {Phys. Rev. Lett.}\ }\textbf {\bibinfo {volume} {87}},\
  \bibinfo {pages} {126804} (\bibinfo {year} {2001})}\BibitemShut {NoStop}%
\bibitem [{\citenamefont {Schreier}\ \emph {et~al.}(2008)\citenamefont
  {Schreier}, \citenamefont {Houck}, \citenamefont {Koch}, \citenamefont
  {Schuster}, \citenamefont {Johnson}, \citenamefont {Chow}, \citenamefont
  {Gambetta}, \citenamefont {Majer}, \citenamefont {Frunzio}, \citenamefont
  {Devoret}, \citenamefont {Girvin},\ and\ \citenamefont
  {Schoelkopf}}]{Schreier_2008}%
  \BibitemOpen
  \bibfield  {author} {\bibinfo {author} {\bibfnamefont {J.~A.}\ \bibnamefont
  {Schreier}}, \bibinfo {author} {\bibfnamefont {A.~A.}\ \bibnamefont {Houck}},
  \bibinfo {author} {\bibfnamefont {J.}~\bibnamefont {Koch}}, \bibinfo {author}
  {\bibfnamefont {D.~I.}\ \bibnamefont {Schuster}}, \bibinfo {author}
  {\bibfnamefont {B.~R.}\ \bibnamefont {Johnson}}, \bibinfo {author}
  {\bibfnamefont {J.~M.}\ \bibnamefont {Chow}}, \bibinfo {author}
  {\bibfnamefont {J.~M.}\ \bibnamefont {Gambetta}}, \bibinfo {author}
  {\bibfnamefont {J.}~\bibnamefont {Majer}}, \bibinfo {author} {\bibfnamefont
  {L.}~\bibnamefont {Frunzio}}, \bibinfo {author} {\bibfnamefont {M.~H.}\
  \bibnamefont {Devoret}}, \bibinfo {author} {\bibfnamefont {S.~M.}\
  \bibnamefont {Girvin}},\ and\ \bibinfo {author} {\bibfnamefont {R.~J.}\
  \bibnamefont {Schoelkopf}},\ }\bibfield  {title} {\bibinfo {title}
  {Suppressing charge noise decoherence in superconducting charge qubits},\
  }\href {https://doi.org/10.1103/PhysRevB.77.180502} {\bibfield  {journal}
  {\bibinfo  {journal} {Phys. Rev. B}\ }\textbf {\bibinfo {volume} {77}},\
  \bibinfo {pages} {180502(R)} (\bibinfo {year} {2008})}\BibitemShut {NoStop}%
\bibitem [{\citenamefont {Bouchiat}\ \emph {et~al.}(1998)\citenamefont
  {Bouchiat}, \citenamefont {Vion}, \citenamefont {Joyez}, \citenamefont
  {Esteve},\ and\ \citenamefont {Devoret}}]{Bouchiat_1998}%
  \BibitemOpen
  \bibfield  {author} {\bibinfo {author} {\bibfnamefont {V.}~\bibnamefont
  {Bouchiat}}, \bibinfo {author} {\bibfnamefont {D.}~\bibnamefont {Vion}},
  \bibinfo {author} {\bibfnamefont {P.}~\bibnamefont {Joyez}}, \bibinfo
  {author} {\bibfnamefont {D.}~\bibnamefont {Esteve}},\ and\ \bibinfo {author}
  {\bibfnamefont {M.~H.}\ \bibnamefont {Devoret}},\ }\bibfield  {title}
  {\bibinfo {title} {Quantum coherence with a single cooper pair},\ }\href
  {https://doi.org/10.1238/Physica.Topical.076a00165} {\bibfield  {journal}
  {\bibinfo  {journal} {Physica Scripta}\ }\textbf {\bibinfo {volume} {1998}},\
  \bibinfo {pages} {165} (\bibinfo {year} {1998})}\BibitemShut {NoStop}%
\bibitem [{\citenamefont {Nakamura}\ \emph {et~al.}(1999)\citenamefont
  {Nakamura}, \citenamefont {Pashkin},\ and\ \citenamefont
  {Tsai}}]{Nakamura_1999}%
  \BibitemOpen
  \bibfield  {author} {\bibinfo {author} {\bibfnamefont {Y.}~\bibnamefont
  {Nakamura}}, \bibinfo {author} {\bibfnamefont {Y.~A.}\ \bibnamefont
  {Pashkin}},\ and\ \bibinfo {author} {\bibfnamefont {J.~S.}\ \bibnamefont
  {Tsai}},\ }\bibfield  {title} {\bibinfo {title} {Coherent control of
  macroscopic quantum states in a single-cooper-pair box},\ }\href
  {https://doi.org/10.1038/19718} {\bibfield  {journal} {\bibinfo  {journal}
  {Nature}\ }\textbf {\bibinfo {volume} {398}},\ \bibinfo {pages} {786}
  (\bibinfo {year} {1999})}\BibitemShut {NoStop}%
\bibitem [{\citenamefont {Anderson}(1970)}]{Anderson_1970}%
  \BibitemOpen
  \bibfield  {author} {\bibinfo {author} {\bibfnamefont {P.~W.}\ \bibnamefont
  {Anderson}},\ }\bibfield  {title} {\bibinfo {title} {A poor man's derivation
  of scaling laws for the kondo problem},\ }\href
  {https://doi.org/10.1088/0022-3719/3/12/008} {\bibfield  {journal} {\bibinfo
  {journal} {Journal of Physics C: Solid State Physics}\ }\textbf {\bibinfo
  {volume} {3}},\ \bibinfo {pages} {2436} (\bibinfo {year} {1970})}\BibitemShut
  {NoStop}%
\bibitem [{\citenamefont {Tsvelick}\ and\ \citenamefont
  {Wiegmann}(1983)}]{Tsvelick_1983}%
  \BibitemOpen
  \bibfield  {author} {\bibinfo {author} {\bibfnamefont {A.}~\bibnamefont
  {Tsvelick}}\ and\ \bibinfo {author} {\bibfnamefont {P.}~\bibnamefont
  {Wiegmann}},\ }\bibfield  {title} {\bibinfo {title} {Exact results in the
  theory of magnetic alloys},\ }\href
  {https://doi.org/10.1080/00018738300101581} {\bibfield  {journal} {\bibinfo
  {journal} {Advances in Physics}\ }\textbf {\bibinfo {volume} {32}},\ \bibinfo
  {pages} {453} (\bibinfo {year} {1983})}\BibitemShut {NoStop}%
\bibitem [{\citenamefont {Landauer}(1976)}]{Landauer_1976}%
  \BibitemOpen
  \bibfield  {author} {\bibinfo {author} {\bibfnamefont {R.}~\bibnamefont
  {Landauer}},\ }\bibfield  {title} {\bibinfo {title} {Can capacitance be
  negative?},\ }\href@noop {} {\bibfield  {journal} {\bibinfo  {journal}
  {Collect. Phenom.}\ }\textbf {\bibinfo {volume} {2}},\ \bibinfo {pages} {167}
  (\bibinfo {year} {1976})}\BibitemShut {NoStop}%
\bibitem [{\citenamefont {Catalan}\ \emph {et~al.}(2015)\citenamefont
  {Catalan}, \citenamefont {Jim{\'e}nez},\ and\ \citenamefont
  {Gruverman}}]{Catalan_2015}%
  \BibitemOpen
  \bibfield  {author} {\bibinfo {author} {\bibfnamefont {G.}~\bibnamefont
  {Catalan}}, \bibinfo {author} {\bibfnamefont {D.}~\bibnamefont
  {Jim{\'e}nez}},\ and\ \bibinfo {author} {\bibfnamefont {A.}~\bibnamefont
  {Gruverman}},\ }\bibfield  {title} {\bibinfo {title} {Negative capacitance
  detected},\ }\href {https://doi.org/10.1038/nmat4195} {\bibfield  {journal}
  {\bibinfo  {journal} {Nature Materials}\ }\textbf {\bibinfo {volume} {14}},\
  \bibinfo {pages} {137} (\bibinfo {year} {2015})}\BibitemShut {NoStop}%
\bibitem [{\citenamefont {Hoffmann}\ \emph {et~al.}(2020)\citenamefont
  {Hoffmann}, \citenamefont {Slesazeck}, \citenamefont {Schroeder},\ and\
  \citenamefont {Mikolajick}}]{Hoffmann_2020}%
  \BibitemOpen
  \bibfield  {author} {\bibinfo {author} {\bibfnamefont {M.}~\bibnamefont
  {Hoffmann}}, \bibinfo {author} {\bibfnamefont {S.}~\bibnamefont {Slesazeck}},
  \bibinfo {author} {\bibfnamefont {U.}~\bibnamefont {Schroeder}},\ and\
  \bibinfo {author} {\bibfnamefont {T.}~\bibnamefont {Mikolajick}},\ }\bibfield
   {title} {\bibinfo {title} {What's next for negative capacitance
  electronics?},\ }\href {https://doi.org/10.1038/s41928-020-00474-9}
  {\bibfield  {journal} {\bibinfo  {journal} {Nature Electronics}\ }\textbf
  {\bibinfo {volume} {3}},\ \bibinfo {pages} {504} (\bibinfo {year}
  {2020})}\BibitemShut {NoStop}%
\bibitem [{\citenamefont {Little}(1964)}]{Little_1964}%
  \BibitemOpen
  \bibfield  {author} {\bibinfo {author} {\bibfnamefont {W.~A.}\ \bibnamefont
  {Little}},\ }\bibfield  {title} {\bibinfo {title} {Possibility of
  synthesizing an organic superconductor},\ }\href
  {https://doi.org/https://doi.org/10.1103/PhysRev.134.A1416} {\bibfield
  {journal} {\bibinfo  {journal} {Phys. Rev.}\ }\textbf {\bibinfo {volume}
  {134}},\ \bibinfo {pages} {A1416} (\bibinfo {year} {1964})}\BibitemShut
  {NoStop}%
\bibitem [{\citenamefont {Hamo}\ \emph {et~al.}(2016)\citenamefont {Hamo},
  \citenamefont {Benyamini}, \citenamefont {Shapir}, \citenamefont {Khivrich},
  \citenamefont {Waissman}, \citenamefont {Kaasbjerg}, \citenamefont {Oreg},
  \citenamefont {von Oppen},\ and\ \citenamefont {Ilani}}]{Hamo_2016}%
  \BibitemOpen
  \bibfield  {author} {\bibinfo {author} {\bibfnamefont {A.}~\bibnamefont
  {Hamo}}, \bibinfo {author} {\bibfnamefont {A.}~\bibnamefont {Benyamini}},
  \bibinfo {author} {\bibfnamefont {I.}~\bibnamefont {Shapir}}, \bibinfo
  {author} {\bibfnamefont {I.}~\bibnamefont {Khivrich}}, \bibinfo {author}
  {\bibfnamefont {J.}~\bibnamefont {Waissman}}, \bibinfo {author}
  {\bibfnamefont {K.}~\bibnamefont {Kaasbjerg}}, \bibinfo {author}
  {\bibfnamefont {Y.}~\bibnamefont {Oreg}}, \bibinfo {author} {\bibfnamefont
  {F.}~\bibnamefont {von Oppen}},\ and\ \bibinfo {author} {\bibfnamefont
  {S.}~\bibnamefont {Ilani}},\ }\bibfield  {title} {\bibinfo {title} {Electron
  attraction mediated by coulomb repulsion},\ }\href
  {https://doi.org/https://doi.org/10.1038/nature18639} {\bibfield  {journal}
  {\bibinfo  {journal} {Nature}\ }\textbf {\bibinfo {volume} {535}},\ \bibinfo
  {pages} {395} (\bibinfo {year} {2016})}\BibitemShut {NoStop}%
\bibitem [{\citenamefont {Placke}\ \emph {et~al.}(2018)\citenamefont {Placke},
  \citenamefont {Pluecker}, \citenamefont {Splettstoesser},\ and\ \citenamefont
  {Wegewijs}}]{Placke_2018}%
  \BibitemOpen
  \bibfield  {author} {\bibinfo {author} {\bibfnamefont {B.~A.}\ \bibnamefont
  {Placke}}, \bibinfo {author} {\bibfnamefont {T.}~\bibnamefont {Pluecker}},
  \bibinfo {author} {\bibfnamefont {J.}~\bibnamefont {Splettstoesser}},\ and\
  \bibinfo {author} {\bibfnamefont {M.~R.}\ \bibnamefont {Wegewijs}},\
  }\bibfield  {title} {\bibinfo {title} {Attractive and driven interactions in
  quantum dots: Mechanisms for geometric pumping},\ }\href
  {https://doi.org/https://doi.org/10.1103/PhysRevB.98.085307} {\bibfield
  {journal} {\bibinfo  {journal} {Phys. Rev. B}\ }\textbf {\bibinfo {volume}
  {98}},\ \bibinfo {pages} {085307} (\bibinfo {year} {2018})}\BibitemShut
  {NoStop}%
\bibitem [{\citenamefont {Herrig}\ \emph {et~al.}(2023)\citenamefont {Herrig},
  \citenamefont {Pixley}, \citenamefont {K{\"o}nig},\ and\ \citenamefont
  {Riwar}}]{Herrig_2023}%
  \BibitemOpen
  \bibfield  {author} {\bibinfo {author} {\bibfnamefont {T.}~\bibnamefont
  {Herrig}}, \bibinfo {author} {\bibfnamefont {J.~H.}\ \bibnamefont {Pixley}},
  \bibinfo {author} {\bibfnamefont {E.~J.}\ \bibnamefont {K{\"o}nig}},\ and\
  \bibinfo {author} {\bibfnamefont {R.~P.}\ \bibnamefont {Riwar}},\ }\bibfield
  {title} {\bibinfo {title} {Quasiperiodic circuit quantum electrodynamics},\
  }\href {https://doi.org/10.1038/s41534-023-00786-6} {\bibfield  {journal}
  {\bibinfo  {journal} {npj Quantum Information}\ }\textbf {\bibinfo {volume}
  {9}},\ \bibinfo {pages} {116} (\bibinfo {year} {2023})}\BibitemShut {NoStop}%
\bibitem [{\citenamefont {Herrig}\ \emph {et~al.}(2024)\citenamefont {Herrig},
  \citenamefont {Koliofoti}, \citenamefont {Pixley}, \citenamefont
  {K{\"o}nig},\ and\ \citenamefont {Riwar}}]{herrig2024}%
  \BibitemOpen
  \bibfield  {author} {\bibinfo {author} {\bibfnamefont {T.}~\bibnamefont
  {Herrig}}, \bibinfo {author} {\bibfnamefont {C.}~\bibnamefont {Koliofoti}},
  \bibinfo {author} {\bibfnamefont {J.~H.}\ \bibnamefont {Pixley}}, \bibinfo
  {author} {\bibfnamefont {E.~J.}\ \bibnamefont {K{\"o}nig}},\ and\ \bibinfo
  {author} {\bibfnamefont {R.-P.}\ \bibnamefont {Riwar}},\ }\href@noop {}
  {\bibinfo {title} {Emulating moir\'e materials with quasiperiodic circuit
  quantum electrodynamics}} (\bibinfo {year} {2024}),\ \Eprint
  {https://arxiv.org/abs/2310.15103} {arXiv:2310.15103 [cond-mat.mes-hall]}
  \BibitemShut {NoStop}%
\bibitem [{\citenamefont {Pustilnik}\ and\ \citenamefont
  {Glazman}(2001)}]{Pustilnik_2001}%
  \BibitemOpen
  \bibfield  {author} {\bibinfo {author} {\bibfnamefont {M.}~\bibnamefont
  {Pustilnik}}\ and\ \bibinfo {author} {\bibfnamefont {L.~I.}\ \bibnamefont
  {Glazman}},\ }\bibfield  {title} {\bibinfo {title} {Kondo effect in real
  quantum dots},\ }\href {https://doi.org/10.1103/PhysRevLett.87.216601}
  {\bibfield  {journal} {\bibinfo  {journal} {Phys. Rev. Lett.}\ }\textbf
  {\bibinfo {volume} {87}},\ \bibinfo {pages} {216601} (\bibinfo {year}
  {2001})}\BibitemShut {NoStop}%
\bibitem [{\citenamefont {Bulla}\ \emph {et~al.}(2008)\citenamefont {Bulla},
  \citenamefont {Costi},\ and\ \citenamefont {Pruschke}}]{Bulla_2008}%
  \BibitemOpen
  \bibfield  {author} {\bibinfo {author} {\bibfnamefont {R.}~\bibnamefont
  {Bulla}}, \bibinfo {author} {\bibfnamefont {T.~A.}\ \bibnamefont {Costi}},\
  and\ \bibinfo {author} {\bibfnamefont {T.}~\bibnamefont {Pruschke}},\
  }\bibfield  {title} {\bibinfo {title} {Numerical renormalization group method
  for quantum impurity systems},\ }\href
  {https://doi.org/10.1103/RevModPhys.80.395} {\bibfield  {journal} {\bibinfo
  {journal} {Rev. Mod. Phys.}\ }\textbf {\bibinfo {volume} {80}},\ \bibinfo
  {pages} {395} (\bibinfo {year} {2008})}\BibitemShut {NoStop}%
\bibitem [{\citenamefont {Hosseinkhani}\ and\ \citenamefont
  {Catelani}(2018)}]{Hosseinkhani_2018}%
  \BibitemOpen
  \bibfield  {author} {\bibinfo {author} {\bibfnamefont {A.}~\bibnamefont
  {Hosseinkhani}}\ and\ \bibinfo {author} {\bibfnamefont {G.}~\bibnamefont
  {Catelani}},\ }\bibfield  {title} {\bibinfo {title} {Proximity effect in
  normal-metal quasiparticle traps},\ }\href
  {https://doi.org/10.1103/PhysRevB.97.054513} {\bibfield  {journal} {\bibinfo
  {journal} {Phys. Rev. B}\ }\textbf {\bibinfo {volume} {97}},\ \bibinfo
  {pages} {054513} (\bibinfo {year} {2018})}\BibitemShut {NoStop}%
\bibitem [{cap()}]{capacitive_footnote}%
  \BibitemOpen
  \href@noop {} {}\bibinfo {note} {Depending on the precise electrostatic
  situation, the electron-electron interactions in the normal metal may be
  modified due to capacitive coupling with the superconductor, see also
  Supplementary Material.}\BibitemShut {Stop}%
\bibitem [{\citenamefont {Gonz\'alez~Rosado}(2021)}]{Gonzalez_2021}%
  \BibitemOpen
  \bibfield  {author} {\bibinfo {author} {\bibfnamefont {L.}~\bibnamefont
  {Gonz\'alez~Rosado}},\ }\emph {\bibinfo {title} {Electron-hole diffusion in
  disordered superconductors}},\ \href@noop {} {Ph.D. thesis},\ \bibinfo
  {school} {RWTH Aachen University} (\bibinfo {year} {2021})\BibitemShut
  {NoStop}%
\bibitem [{\citenamefont {Leggett}(1984)}]{Leggett1984}%
  \BibitemOpen
  \bibfield  {author} {\bibinfo {author} {\bibfnamefont {A.~J.}\ \bibnamefont
  {Leggett}},\ }\bibfield  {title} {\bibinfo {title} {Quantum tunneling in the
  presence of an arbitrary linear dissipation mechanism},\ }\href
  {https://doi.org/10.1103/PhysRevB.30.1208} {\bibfield  {journal} {\bibinfo
  {journal} {Phys. Rev. B}\ }\textbf {\bibinfo {volume} {30}},\ \bibinfo
  {pages} {1208} (\bibinfo {year} {1984})}\BibitemShut {NoStop}%
\bibitem [{\citenamefont {Ankerhold}\ and\ \citenamefont
  {Pollak}(2007)}]{Ankerhold2007}%
  \BibitemOpen
  \bibfield  {author} {\bibinfo {author} {\bibfnamefont {J.}~\bibnamefont
  {Ankerhold}}\ and\ \bibinfo {author} {\bibfnamefont {E.}~\bibnamefont
  {Pollak}},\ }\bibfield  {title} {\bibinfo {title} {Dissipation can enhance
  quantum effects},\ }\href {https://doi.org/10.1103/PhysRevE.75.041103}
  {\bibfield  {journal} {\bibinfo  {journal} {Phys. Rev. E}\ }\textbf {\bibinfo
  {volume} {75}},\ \bibinfo {pages} {041103} (\bibinfo {year}
  {2007})}\BibitemShut {NoStop}%
\bibitem [{\citenamefont {Ulrich}\ and\ \citenamefont
  {Hassler}(2016)}]{Ulrich_2016}%
  \BibitemOpen
  \bibfield  {author} {\bibinfo {author} {\bibfnamefont {J.}~\bibnamefont
  {Ulrich}}\ and\ \bibinfo {author} {\bibfnamefont {F.}~\bibnamefont
  {Hassler}},\ }\bibfield  {title} {\bibinfo {title} {Dual approach to circuit
  quantization using loop charges},\ }\href
  {https://doi.org/10.1103/PhysRevB.94.094505} {\bibfield  {journal} {\bibinfo
  {journal} {Phys. Rev. B}\ }\textbf {\bibinfo {volume} {94}},\ \bibinfo
  {pages} {094505} (\bibinfo {year} {2016})}\BibitemShut {NoStop}%
\bibitem [{\citenamefont {Kaur}\ \emph {et~al.}(2021)\citenamefont {Kaur},
  \citenamefont {S\'epulcre}, \citenamefont {Roch}, \citenamefont {Snyman},
  \citenamefont {Florens},\ and\ \citenamefont {Bera}}]{Kaur_2021}%
  \BibitemOpen
  \bibfield  {author} {\bibinfo {author} {\bibfnamefont {K.}~\bibnamefont
  {Kaur}}, \bibinfo {author} {\bibfnamefont {T.}~\bibnamefont {S\'epulcre}},
  \bibinfo {author} {\bibfnamefont {N.}~\bibnamefont {Roch}}, \bibinfo {author}
  {\bibfnamefont {I.}~\bibnamefont {Snyman}}, \bibinfo {author} {\bibfnamefont
  {S.}~\bibnamefont {Florens}},\ and\ \bibinfo {author} {\bibfnamefont
  {S.}~\bibnamefont {Bera}},\ }\bibfield  {title} {\bibinfo {title} {Spin-boson
  quantum phase transition in multilevel superconducting qubits},\ }\href
  {https://doi.org/https://doi.org/10.1103/PhysRevLett.127.237702} {\bibfield
  {journal} {\bibinfo  {journal} {Phys. Rev. Lett.}\ }\textbf {\bibinfo
  {volume} {127}},\ \bibinfo {pages} {237702} (\bibinfo {year}
  {2021})}\BibitemShut {NoStop}%
\bibitem [{\citenamefont {Ambegaokar}\ \emph {et~al.}(1982)\citenamefont
  {Ambegaokar}, \citenamefont {Eckern},\ and\ \citenamefont
  {Sch\"on}}]{Ambegaokar_1982}%
  \BibitemOpen
  \bibfield  {author} {\bibinfo {author} {\bibfnamefont {V.}~\bibnamefont
  {Ambegaokar}}, \bibinfo {author} {\bibfnamefont {U.}~\bibnamefont {Eckern}},\
  and\ \bibinfo {author} {\bibfnamefont {G.}~\bibnamefont {Sch\"on}},\
  }\bibfield  {title} {\bibinfo {title} {Quantum dynamics of tunneling between
  superconductors},\ }\href {https://doi.org/10.1103/PhysRevLett.48.1745}
  {\bibfield  {journal} {\bibinfo  {journal} {Phys. Rev. Lett.}\ }\textbf
  {\bibinfo {volume} {48}},\ \bibinfo {pages} {1745} (\bibinfo {year}
  {1982})}\BibitemShut {NoStop}%
\bibitem [{\citenamefont {Eckern}\ \emph {et~al.}(1984)\citenamefont {Eckern},
  \citenamefont {Sch\"on},\ and\ \citenamefont {Ambegaokar}}]{Ambegaokar_1984}%
  \BibitemOpen
  \bibfield  {author} {\bibinfo {author} {\bibfnamefont {U.}~\bibnamefont
  {Eckern}}, \bibinfo {author} {\bibfnamefont {G.}~\bibnamefont {Sch\"on}},\
  and\ \bibinfo {author} {\bibfnamefont {V.}~\bibnamefont {Ambegaokar}},\
  }\bibfield  {title} {\bibinfo {title} {Quantum dynamics of a superconducting
  tunnel junction},\ }\href {https://doi.org/10.1103/PhysRevB.30.6419}
  {\bibfield  {journal} {\bibinfo  {journal} {Phys. Rev. B}\ }\textbf {\bibinfo
  {volume} {30}},\ \bibinfo {pages} {6419} (\bibinfo {year}
  {1984})}\BibitemShut {NoStop}%
\bibitem [{Note3()}]{Note3}%
  \BibitemOpen
  \bibinfo {note} {The word transmon is commonly associated with very high
  ratios $E_J/E_C$ (of order $\sim 50$ or higher). We here use the term a
  little more flexibly, and mean simply that $E_J$ shall be sufficiently large
  compared to $E_C$, such that $\protect \sqrt {E_J E_C}>E_S$.}\BibitemShut
  {Stop}%
\bibitem [{\citenamefont {Rymarz}\ \emph {et~al.}(2021)\citenamefont {Rymarz},
  \citenamefont {Bosco}, \citenamefont {Ciani},\ and\ \citenamefont
  {DiVincenzo}}]{Rymarz2021}%
  \BibitemOpen
  \bibfield  {author} {\bibinfo {author} {\bibfnamefont {M.}~\bibnamefont
  {Rymarz}}, \bibinfo {author} {\bibfnamefont {S.}~\bibnamefont {Bosco}},
  \bibinfo {author} {\bibfnamefont {A.}~\bibnamefont {Ciani}},\ and\ \bibinfo
  {author} {\bibfnamefont {D.~P.}\ \bibnamefont {DiVincenzo}},\ }\bibfield
  {title} {\bibinfo {title} {Hardware-encoding grid states in a nonreciprocal
  superconducting circuit},\ }\href
  {https://doi.org/https://doi.org/10.1103/PhysRevX.11.011032} {\bibfield
  {journal} {\bibinfo  {journal} {Phys. Rev. X}\ }\textbf {\bibinfo {volume}
  {11}},\ \bibinfo {pages} {011032} (\bibinfo {year} {2021})}\BibitemShut
  {NoStop}%
\bibitem [{\citenamefont {Maile}\ \emph {et~al.}(2018)\citenamefont {Maile},
  \citenamefont {Andergassen}, \citenamefont {Belzig},\ and\ \citenamefont
  {Rastelli}}]{Maile2018}%
  \BibitemOpen
  \bibfield  {author} {\bibinfo {author} {\bibfnamefont {D.}~\bibnamefont
  {Maile}}, \bibinfo {author} {\bibfnamefont {S.}~\bibnamefont {Andergassen}},
  \bibinfo {author} {\bibfnamefont {W.}~\bibnamefont {Belzig}},\ and\ \bibinfo
  {author} {\bibfnamefont {G.}~\bibnamefont {Rastelli}},\ }\bibfield  {title}
  {\bibinfo {title} {Quantum phase transition with dissipative frustration},\
  }\href {https://doi.org/10.1103/PhysRevB.97.155427} {\bibfield  {journal}
  {\bibinfo  {journal} {Phys. Rev. B}\ }\textbf {\bibinfo {volume} {97}},\
  \bibinfo {pages} {155427} (\bibinfo {year} {2018})}\BibitemShut {NoStop}%
\bibitem [{\citenamefont {Maile}\ \emph {et~al.}(2022)\citenamefont {Maile},
  \citenamefont {Ankerhold}, \citenamefont {Andergassen}, \citenamefont
  {Belzig},\ and\ \citenamefont {Rastelli}}]{Maile2022}%
  \BibitemOpen
  \bibfield  {author} {\bibinfo {author} {\bibfnamefont {D.}~\bibnamefont
  {Maile}}, \bibinfo {author} {\bibfnamefont {J.}~\bibnamefont {Ankerhold}},
  \bibinfo {author} {\bibfnamefont {S.}~\bibnamefont {Andergassen}}, \bibinfo
  {author} {\bibfnamefont {W.}~\bibnamefont {Belzig}},\ and\ \bibinfo {author}
  {\bibfnamefont {G.}~\bibnamefont {Rastelli}},\ }\bibfield  {title} {\bibinfo
  {title} {Engineering the speedup of quantum tunneling in josephson systems
  via dissipation},\ }\href {https://doi.org/10.1103/PhysRevB.106.045408}
  {\bibfield  {journal} {\bibinfo  {journal} {Phys. Rev. B}\ }\textbf {\bibinfo
  {volume} {106}},\ \bibinfo {pages} {045408} (\bibinfo {year}
  {2022})}\BibitemShut {NoStop}%
\bibitem [{\citenamefont {Abanov}\ \emph {et~al.}(2011)\citenamefont {Abanov},
  \citenamefont {Ivanov},\ and\ \citenamefont {Qian}}]{Abanov2011}%
  \BibitemOpen
  \bibfield  {author} {\bibinfo {author} {\bibfnamefont {A.~G.}\ \bibnamefont
  {Abanov}}, \bibinfo {author} {\bibfnamefont {D.~A.}\ \bibnamefont {Ivanov}},\
  and\ \bibinfo {author} {\bibfnamefont {Y.}~\bibnamefont {Qian}},\ }\bibfield
  {title} {\bibinfo {title} {Quantum fluctuations of one-dimensional free
  fermions and {F}isher--{H}artwig formula for {T}oeplitz determinants},\
  }\href {https://doi.org/10.1088/1751-8113/44/48/485001} {\bibfield  {journal}
  {\bibinfo  {journal} {J. Phys. A Math. Theor.}\ }\textbf {\bibinfo {volume}
  {44}},\ \bibinfo {pages} {485001} (\bibinfo {year} {2011})}\BibitemShut
  {NoStop}%
\bibitem [{\citenamefont {Song}\ \emph {et~al.}(2012)\citenamefont {Song},
  \citenamefont {Rachel}, \citenamefont {Flindt}, \citenamefont {Klich},
  \citenamefont {Laflorencie},\ and\ \citenamefont {Le~Hur}}]{LeHur2012}%
  \BibitemOpen
  \bibfield  {author} {\bibinfo {author} {\bibfnamefont {H.~F.}\ \bibnamefont
  {Song}}, \bibinfo {author} {\bibfnamefont {S.}~\bibnamefont {Rachel}},
  \bibinfo {author} {\bibfnamefont {C.}~\bibnamefont {Flindt}}, \bibinfo
  {author} {\bibfnamefont {I.}~\bibnamefont {Klich}}, \bibinfo {author}
  {\bibfnamefont {N.}~\bibnamefont {Laflorencie}},\ and\ \bibinfo {author}
  {\bibfnamefont {K.}~\bibnamefont {Le~Hur}},\ }\bibfield  {title} {\bibinfo
  {title} {Bipartite fluctuations as a probe of many-body entanglement},\
  }\href {https://doi.org/10.1103/PhysRevB.85.035409} {\bibfield  {journal}
  {\bibinfo  {journal} {Phys. Rev. B}\ }\textbf {\bibinfo {volume} {85}},\
  \bibinfo {pages} {035409} (\bibinfo {year} {2012})}\BibitemShut {NoStop}%
\bibitem [{\citenamefont {Houzet}\ and\ \citenamefont
  {Glazman}(2019)}]{Houzet_2019}%
  \BibitemOpen
  \bibfield  {author} {\bibinfo {author} {\bibfnamefont {M.}~\bibnamefont
  {Houzet}}\ and\ \bibinfo {author} {\bibfnamefont {L.~I.}\ \bibnamefont
  {Glazman}},\ }\bibfield  {title} {\bibinfo {title} {Microwave spectroscopy of
  a weakly pinned charge density wave in a superinductor},\ }\href
  {https://doi.org/https://doi.org/10.1103/PhysRevLett.122.237701} {\bibfield
  {journal} {\bibinfo  {journal} {Phys. Rev. Lett.}\ }\textbf {\bibinfo
  {volume} {122}},\ \bibinfo {pages} {237701} (\bibinfo {year}
  {2019})}\BibitemShut {NoStop}%
\bibitem [{\citenamefont {Hekking}\ and\ \citenamefont
  {Glazman}(1997)}]{Hekking_1997}%
  \BibitemOpen
  \bibfield  {author} {\bibinfo {author} {\bibfnamefont {F.~W.~J.}\
  \bibnamefont {Hekking}}\ and\ \bibinfo {author} {\bibfnamefont {L.~I.}\
  \bibnamefont {Glazman}},\ }\bibfield  {title} {\bibinfo {title} {Quantum
  fluctuations in the equilibrium state of a thin superconducting loop},\
  }\href {https://doi.org/10.1103/PhysRevB.55.6551} {\bibfield  {journal}
  {\bibinfo  {journal} {Phys. Rev. B}\ }\textbf {\bibinfo {volume} {55}},\
  \bibinfo {pages} {6551} (\bibinfo {year} {1997})}\BibitemShut {NoStop}%
\bibitem [{\citenamefont {Pop}\ \emph {et~al.}(2012)\citenamefont {Pop},
  \citenamefont {Dou\ifmmode~\mbox{\c{c}}\else \c{c}\fi{}ot}, \citenamefont
  {Ioffe}, \citenamefont {Protopopov}, \citenamefont {Lecocq}, \citenamefont
  {Matei}, \citenamefont {Buisson},\ and\ \citenamefont {Guichard}}]{Pop2012}%
  \BibitemOpen
  \bibfield  {author} {\bibinfo {author} {\bibfnamefont {I.~M.}\ \bibnamefont
  {Pop}}, \bibinfo {author} {\bibfnamefont {B.}~\bibnamefont
  {Dou\ifmmode~\mbox{\c{c}}\else \c{c}\fi{}ot}}, \bibinfo {author}
  {\bibfnamefont {L.}~\bibnamefont {Ioffe}}, \bibinfo {author} {\bibfnamefont
  {I.}~\bibnamefont {Protopopov}}, \bibinfo {author} {\bibfnamefont
  {F.}~\bibnamefont {Lecocq}}, \bibinfo {author} {\bibfnamefont
  {I.}~\bibnamefont {Matei}}, \bibinfo {author} {\bibfnamefont
  {O.}~\bibnamefont {Buisson}},\ and\ \bibinfo {author} {\bibfnamefont
  {W.}~\bibnamefont {Guichard}},\ }\bibfield  {title} {\bibinfo {title}
  {Experimental demonstration of aharonov-casher interference in a josephson
  junction circuit},\ }\href {https://doi.org/10.1103/PhysRevB.85.094503}
  {\bibfield  {journal} {\bibinfo  {journal} {Phys. Rev. B}\ }\textbf {\bibinfo
  {volume} {85}},\ \bibinfo {pages} {094503} (\bibinfo {year}
  {2012})}\BibitemShut {NoStop}%
\bibitem [{\citenamefont {Astafiev}\ \emph {et~al.}(2012)\citenamefont
  {Astafiev}, \citenamefont {Ioffe}, \citenamefont {Kafanov}, \citenamefont
  {Pashkin}, \citenamefont {Arutyunov}, \citenamefont {Shahar}, \citenamefont
  {Cohen},\ and\ \citenamefont {Tsai}}]{Astafiev_2012}%
  \BibitemOpen
  \bibfield  {author} {\bibinfo {author} {\bibfnamefont {O.~V.}\ \bibnamefont
  {Astafiev}}, \bibinfo {author} {\bibfnamefont {L.~B.}\ \bibnamefont {Ioffe}},
  \bibinfo {author} {\bibfnamefont {S.}~\bibnamefont {Kafanov}}, \bibinfo
  {author} {\bibfnamefont {Y.~A.}\ \bibnamefont {Pashkin}}, \bibinfo {author}
  {\bibfnamefont {K.~Y.}\ \bibnamefont {Arutyunov}}, \bibinfo {author}
  {\bibfnamefont {D.}~\bibnamefont {Shahar}}, \bibinfo {author} {\bibfnamefont
  {O.}~\bibnamefont {Cohen}},\ and\ \bibinfo {author} {\bibfnamefont {J.~S.}\
  \bibnamefont {Tsai}},\ }\bibfield  {title} {\bibinfo {title} {Coherent
  quantum phase slip},\ }\href
  {https://doi.org/https://doi.org/10.1038/nature10930} {\bibfield  {journal}
  {\bibinfo  {journal} {Nature}\ }\textbf {\bibinfo {volume} {484}},\ \bibinfo
  {pages} {355} (\bibinfo {year} {2012})}\BibitemShut {NoStop}%
\end{thebibliography}%

\end{document}